\documentclass[fleqn,usenatbib]{mnras}

\usepackage{newtxtext,newtxmath}
\usepackage[T1]{fontenc}

\DeclareRobustCommand{\VAN}[3]{#2}
\let\VANthebibliography\thebibliography
\def\thebibliography{\DeclareRobustCommand{\VAN}[3]{##3}\VANthebibliography}

\usepackage{graphicx}	% Including figure files
\usepackage{amsmath}	% Advanced maths commands
\usepackage{amssymb}	% Extra maths symbols

\usepackage[utf8]{inputenc}
\usepackage{graphicx}

\graphicspath{{figures/}}
\usepackage{epsf}
\usepackage{soul}
\usepackage{tabularx}
\usepackage{xspace}
\usepackage{aas_macros}
\usepackage{array}
\usepackage{url}
\usepackage{amsmath}
\usepackage{amssymb}
\usepackage[noabbrev, capitalise]{cleveref}
\usepackage{color,graphicx}
\usepackage{appendix}
\usepackage{verbatim}

\usepackage[dvipsnames]{xcolor}

\def\simlt{\mathrel{\rlap{\lower 3pt\hbox{$\sim$}}\raise 2.0pt\hbox{$<$}}}
\def\simgt{\mathrel{\rlap{\lower 3pt\hbox{$\sim$}} \raise
2.0pt\hbox{$>$}}}

\title[SKA Science Data Challenge 1: analysis and results]{Square Kilometre Array Science Data Challenge 1: analysis and results}

\author[Bonaldi et al.]{A. Bonaldi,$^{1,2}\thanks{a.bonaldi@skatelescope.org}$
T. An$^3$, %shanghai
M. Br\"uggen$^{4}$, %hs
S. Burkutean$^{5}$, %CACAO
B. Coelho$^{6}$, %engageSKA-P.
H. Goodarzi$^7$, %IPM
\newauthor
P. Hartley$^{1}$, %SKAO
P.~K. Sandhu$^{8}$, %RADGK
C. Wu$^{9}$, %ICRAR
L. Yu$^{10}$, %NAOC
M.~H. Zhoolideh Haghighi$^7$, %IPM2
\newauthor
S. Antón$^{11,6}$,
Z. Bagheri$^{7,12}$, %IPM
D. Barbosa$^6$,
J.~P. Barraca$^{6,13}$,
D. Bartashevich$^6$,
\newauthor
M. Bergano$^6$,
M. Bonato$^{5}$,
J. Brand$^{5}$,
F. de Gasperin$^{4}$,
A. Giannetti$^{5}$,
R. Dodson$^{9}$, 
\newauthor
P. Jain$^{8}$,
S. Jaiswal$^3$, 
B. Lao$^3$,
B. Liu$^{10}$,
E. Liuzzo$^{5}$,
Y. Lu$^3$,
V. Lukic$^{4}$,
D. Maia$^{14}$,
\newauthor
N. Marchili$^{5}$,
M. Massardi$^{5}$,
P. Mohan$^3$, 
J.~B. Morgado$^{14}$,
M. Panwar$^{8}$,
Prabhakar$^{8}$,
\newauthor
V.~A.~R.~M. Ribeiro$^{6,15}$,
K.~L.~J. Rygl$^{5}$,
V. Sabz Ali$^7$, 
E. Saremi$^7$,
E. Schisano$^{16}$,
\newauthor
S. Sheikhnezami$^{17,7}$,
A. Vafaei Sadr $^{18}$
A. Wong$^{19}$,
O.~I. Wong$^{9,21,20}$
\\
Affiliations are at the end of the paper\\
}

\date{Accepted XXX. Received YYY; in original form ZZZ}

\pubyear{2020}

\begin{document}
\label{firstpage}
\maketitle

\begin{abstract}
As the largest radio telescope in the world, the Square Kilometre Array (SKA) will lead
the next generation of radio astronomy. The feats of engineering required to construct the
telescope array will be matched only by the techniques developed to exploit the rich scientific
value of the data. To drive forward the development of efficient and accurate analysis methods,
we are designing a series of data challenges that will provide the scientific community with
high-quality datasets for testing and evaluating new techniques. In this paper we present a
description and results from the first such Science Data Challenge (SDC1). Based on SKA
MID continuum simulated observations and covering three frequencies (560 MHz, 1400MHz
and 9200 MHz) at three depths (8\,h, 100\,h and 1000\,h), SDC1 asked participants to apply
source detection, characterization and classification methods to simulated data. The challenge
opened in November 2018, with nine teams submitting results by the deadline of April 2019. In
this work we analyse the results for 8 of those teams, showcasing the variety of approaches
that can be successfully used to find, characterise and classify sources in a deep, crowded field.
The results also demonstrate the importance of building domain knowledge and expertise on
this kind of analysis to obtain the best performance. As high-resolution
observations begin revealing the true complexity of the sky, one of the outstanding challenges
emerging from this analysis is the ability to deal with highly resolved and complex sources as
effectively as the unresolved source population.
\end{abstract}

%\date{October 2019}

%\begin{document}

%\maketitle

\section{Introduction}\label{sec:intro}

The Square Kilometre Array (SKA,\footnote{\url{https://skatelescope.org}}) will be the world's largest radio telescope. 
The sensitivity and image quality of the SKA define new opportunities for science exploitation but also new challenges for data analysis.  The overwhelming volume of raw SKA data means that they typically cannot be delivered to the Principal Investigators (PIs) or to the Key Science Project (KSP) teams, who instead will have access to calibrated and gridded imaging/non imaging products, called SKA Observatory (Observatory, in short hereafter) data products. Thus, part of the analysis that, on currently operating radio facilities, research groups typically carry out themselves will be the responsibility of the Observatory. For this change in handover point to be successful, it is necessary to build a good understanding of the needs of the science community and of the nature of the Observatory data products.  

One of the actions that has been implemented in order to support the necessary development is the SKA Science Data Challenges (SDCs). These challenges, which are being regularly issued to the community, may consist of real data from currently operating radio facilities or of simulated SKA data. To make each challenge manageable and provide a training opportunity, the full SKA complexity is reached in steps. The goal of each SDC is to exercise some aspects of the analysis that will ultimately be performed on an Observatory data product.

In this paper we discuss the aims and the outcome of the SKA Science Data Challenge \#1 \citep[SDC1,][]{2018arXiv181110454B}. 
SDC1 addresses source finding, characterization and classification for radio continuum sources, on SKA MID simulated images. Many source finding methods have been developed and tested in the literature \citep[e.g.][]{1996A&AS..117..393B,2002AJ....123.1086H,2011A&A...530A.133M,2014AIPC.1636...55F,2012MNRAS.425..979H,2015ascl.soft02007M,2012PASA...29..371W,2012MNRAS.421.3242W,2018PASA...35...11H,Hale2019,2019Galax...8....3L,wu19}.
Comparisons of performance of some of these methods in the context of ASKAP and VLA observations are presented in \cite{hancock2012}, \cite{tessa2016}  and  \cite{hopkins2015}. Rather than performing a similar analysis in the context of the SKA, the aim of SDC1 is  to facilitate further work on the subject and encourage a larger participation to this field, which in time can lead to new ideas and new methods for the future SKA surveys. Another goal is to familiarise the community with the complexity of the SKA data and the challenges posed by its analysis, and to provide a training opportunity to overcome them using the SDC1 dataset as a concrete example. 

In line with our goal of community engagement, participation in SDC1 was not restricted to developers of source-finding methods or their most expert users, but instead open to any team that wanted to engage in it. Teams could use their own combination of publicly available and purpose-developed software, with no requirement for the methods to be independent from one-another or their list to be exhaustive. In line with our goal of exposing the future SKA challenges, SDC1 constitutes a significant step forward in the complexity of the dataset with respect to previous work, in terms of sheer number of sources (of the order of $10^5$ per square degree, down to well below the 1\,$\sigma$ noise levels), and  source properties (we include resolved sources and multi-component sources with complex morphology). 

While these aspects make SDC1 not straightforward in terms of direct comparison of method performance, they give more of a global view of the outstanding challenges at the scale and complexity of the SKA, and of the preparedness of the community.  By delivering the dataset publicly \footnote{\url{ https://astronomers.skatelescope.org/ska-science-data-challenge-1/}}  as well as the software to evaluate one own's performance \footnote{\url{https://pypi.org/project/ska-sdc/}} , we leave open an opportunity for the training to continue and the results to improve beyond what is presented in this paper.

The outline of the paper is as follows: Sec. \ref{sec:sdc1} describes the SDC1 dataset and the challenge; Sec. \ref{sec:methods} describes the teams that participated and the methods they used; Sec. \ref{sec:evaluation} explains how the submissions were evaluated and scored; Sec. \ref{sec:results} shows the results and finally Sec. \ref{sec:conclusions} presents our conclusions.

\section{SDC1 definition} \label{sec:sdc1}
\subsection{The dataset}
The SDC1 dataset was released on the 25th of November 2019 and it is available on the SKA astronomers website  \url{https://astronomers.skatelescope.org/ska-science-data-challenge-1/}. It consists of 9 image files, in FITS format. Each file is a simulated SKA continuum image in total intensity at 3 frequencies:
\begin{enumerate}
\item	560\,MHz, representative of SKA Mid Band 1
\item	1.4\,GHz, representative of SKA Mid Band 2
\item	9.2\,GHz, representative of SKA Mid Band 5
\end{enumerate}
Furthermore, 3 telescope integration depths per frequency are provided: 
\begin{enumerate}
\item	8\,h, representative of a single-track observation;
\item	100\,h, representative of a medium-depth integration;
\item	1000\,h, representative of a deep integration;
\end{enumerate}

The simulated field is nominally centred at RA=0, Dec=$-$30 for each map. The sky model is a plausible realization of the radio sky at those frequencies, but there is no attempt to make it similar to the actual sky at those coordinates. The nine maps share the same sky model realizations, to allow cross-matching between frequencies and direct comparisons between results for different noise levels. 

The simulated observation strategy is that of a single telescope pointing. This means that the sensitivity of the array is maximum at the centre coordinates and decreases towards its outskirts, as described by the primary beam. 
The Field of View (FoV) was chosen for each frequency to contain the primary beam out to the first null. This gives a map size of 5.5, 2.2 and 0.33 degrees on a side for 560\,MHz, 1.4\,GHz and 9.2\,GHz respectively. 

The number of pixels on a side is always 32,768, which gives a pixel size of 0.60, 0.24, and 0.037\,arcsec for 560\,MHz, 1.4\,GHz and 9.2\,GHz, respectively. The imaging resolution in the Gaussian approximation is 1.5, 0.6 and 0.09\,arcsec FWHM for 560\,MHz, 1.4\,GHz and 9.2\,GHz respectively. A more accurate description of the resolution (including sidelobes) is given by the synthesized beam.  Both the primary beam and the synthesized beam are provided as ancillary data. 
Along with the images, a catalogue revealing only a fraction of the simulated galaxies was released for each frequency band. This aims to test and debug the procedures, as well as providing a training set for methods requiring training. This catalogue lists all of the embedded sources within 5\% of the FoV area and unveils their properties. 
Full details of how the dataset was generated can be found in \cite{2018arXiv181110454B}; below, we provide a summary of the steps taken.
The simulated dataset was produced by first using the Tiered Radio Extragalactic Continuum Simulation software (T-RECS)\footnote{\url{https://github.com/abonaldi/TRECS}} \citep{2019MNRAS.482....2B} to generate a sky model catalogue containing star-forming galaxies (SFGs) and Active Galactic Nuclei (AGN),  with integrated flux densities, sky coordinates, and size and shape information attributed to each source. For SFGs, a redshift-dependent luminosity function is generated by exploiting the tight correlation with star-formation rate (SFR), the evolution of which is well studied. For AGN, an evolutionary luminosity function model representing steep-spectrum sources (SS-AGN), flat-spectrum radio quasars (FSRQs) and BL Lac was adopted. 

\begin{figure}
\includegraphics[width=7.5cm]{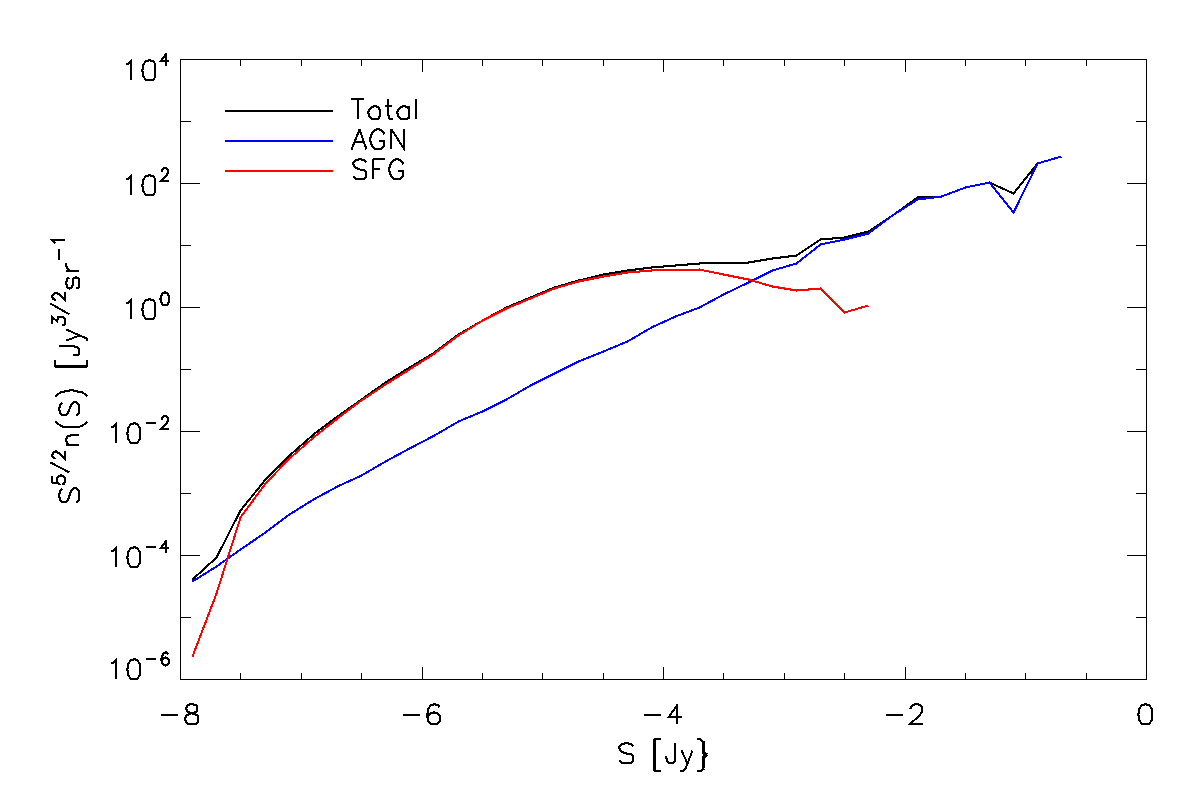}

\caption{Differential source counts of the SDC1 simulation at 1.4\,GHz.}
\label{fig:diffcounts}

\end{figure}

Figure \ref{fig:diffcounts} shows the differential source counts of the simulation at 1.4\,GHz. Agreement of the counts with observations, as well as luminosity functions and redshift distributions, are shown in \cite{2019MNRAS.482....2B}. Radio-loud AGN dominate the counts above the mJy level; the simulation is complete above an integrated flux of $10^{-7.5}$\,Jy.

A morphological model representing the properties of each source was then injected onto the simulated field.
The size of each source was used to determine whether it would appear to be `resolved' or `unresolved' on the sky at each observing frequency, with a cut made at a size $\geq$ or $<$ of 3 pixels, respectively. Images representing the morphological structure of the extended SFG population were generated using a Galsim- \citep{2015A&C....10..121R} based pipeline developed for the SuperCLASS project\footnote{\url{http://www.e-merlin.ac.uk/legacy/projects/superclass.html}} \citep{2020arXiv200301736H}. Each SFG was modelled using an exponential Sersic profile, projected into an ellipsoid with a given axis ratio and position angle. Extended SS-AGN sources were created from a library of real, high-resolution images to which a set of scaling, rotation and reflection transformations were applied. We note that some subclasses of morphologies, such as giant radio galaxies or wide-angle tailed galaxies, could be under-represented in the image library and therefore in the simulation. Resolved flat-spectrum sources were modelled using Gaussian and point component pairs. Once generated, images of extended sources were added  as ``postage stamps'' to the full sky field. Compact sources from all populations were added to the image as elliptical Gaussian components.

Once added to the sky, all sources underwent a convolution using a FWHM of two pixels in order to produce a gridded sky model, before a primary beam attenuation was applied. Visibility files were generated using the locations of all 197 dishes of the SKA1-Mid and MeerKAT configurations for the 1400~MHz band, while the 560 and 9200~MHz bands used the 133 SKA1-Mid locations\footnote{The overlap between the MeerKAT UHF band, 580--1050\,MHz, and SDC1 Band1, 480--650\,MHz, is only partial}. Time and frequency sampling was significantly reduced with respect to full operational sampling rates, but was based on a 30\% fractional bandwidth in frequency and hour angle coverage spanning -4 to +4 hours of Local Sidereal Time. An FFT-based convolution of the natural visibility density grid was used to calculate so-called "uniform" gridding weights, before application of a Gaussian taper chosen to match the target FWHM of 1.5, 0.60 and 0.09 arcsec in the 560, 1400 and 9200~MHz bands, respectively.

Due to the overwhelming volume of raw data, the SKA data reduction workflow is such that calibration and imaging happen on single-track observations, generating Observatory Data Products that are accumulated and co-added to  produce  deeper images. To  reflect  this, deconvolution has been simulated for all our images down to a depth appropriate for an 8\,h observation. 
Image deconvolution effects were simulated by applying a clipping threshold (three times the expected RMS noise level of an 8 hour observation) to the primary beam tapered sky model, the brightness distribution above the threshold was convolved with a Gaussian restoring beam -- taking explicit account of the sky model gridding convolution function -- and below this threshold was convolved with the relevant dirty beam (again taking explicit account of the gridding convolution). 
The sum of these two images represents the simulated observed sky model.

In an interferometric observation, both the noise and the sky signal are processed through the same spatial frequency filter.  To get noise fields that are an accurate representation of the expected correlated image noise, 
dirty noise images were generated from the same imaging simulations described above to produce the synthesized beams.
The RMS noise amplitude was scaled to represent values appropriate for the three different simulation depths and frequency bands under consideration. Different noise realisations were used in all cases, to ensure that the noise would not correlate amongst either image depth or between frequency bands.
The RMS noise levels are reported in Table \ref{tab:noise}.

For comparison, the last row of Table \ref{tab:noise} also reports the classical confusion limit, computed as a function of the beam size $\theta$ and frequency $\nu$ following \cite{2012ApJ...758...23C}
\begin{equation}
\sigma_{\rm c}=1.2 \mu {\rm Jy\,beam}^{-1}\left( \frac{\nu}{3.02\,{\rm GHz}}\right) ^{-0.7} \left(\frac{\theta}{8''}\right)^{10/3},
\end{equation}
which shows that the maps are still noise-limited even at the deepest exposure.

\begin{table}
    \centering
    \caption{Noise RMS [nJy/beam] of the simulated maps per frequency and exposure, compared to the classical confusion limit.  \label{tab:noise}}
    \begin{tabular}{c|c|c|c|}
&560\,MHz&1400\,MHz&9200\,MHz  \\
\hline
8\,h&2880&710&430\\
100\,h&810&200&120\\
1000\,h&255&73&38\\
\hline
confusion&15&0.36&0.0002\\%confusion limits -check beam areas. 
 \end{tabular}

\end{table}
The final data products were the sum of the simulated sky model and the relevant noise image described above. It is worth mentioning that the image quality is unrealistically good, because no systematic effects such as calibration errors, pointing errors or deconvolution errors were injected \citep[see][for more details]{2018arXiv181110454B}.

\begin{figure}
\includegraphics[width=7.5cm]{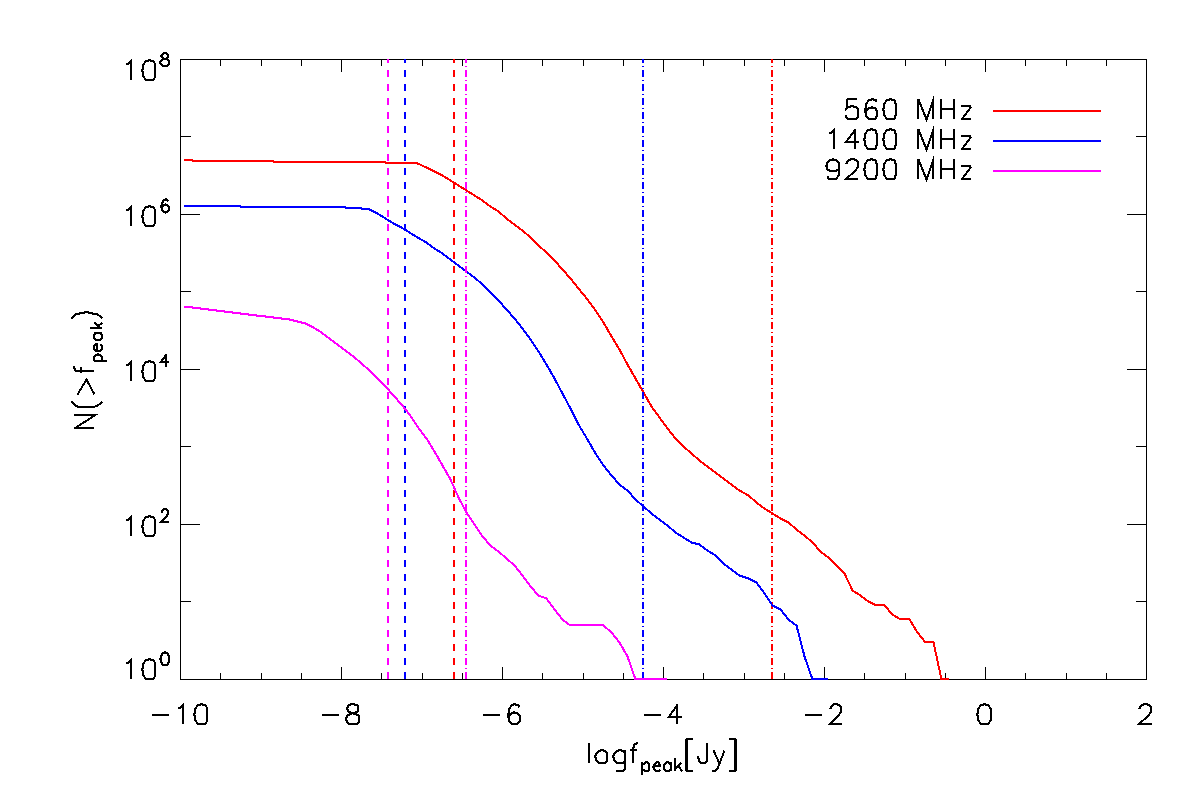}

\caption{Cumulative source counts for the three simulated maps for the full FoV as a function of apparent peak flux. The vertical dashed lines are the 1$\sigma$ noise levels for the 1000\,h exposure.}
\label{fig:intcounts}

\end{figure}

Figure \ref{fig:intcounts} shows the cumulative counts for the sources that were injected in the maps at the three considered frequencies. The counts are shown as a function of the peak apparent flux; they have not been normalised for the different sky areas, therefore they directly give the total number of sources in the maps for various flux thresholds.  

The flattening of the curves for the lowest fluxes is artificial, and is due to a flux threshold applied to the sources for computational reasons. However, 
 the injection of sources goes well below the  1\,$\sigma$ noise levels for the deepest exposure, which is indicated by the dashed vertical lines. This means that our dataset contains a background noise due to fainter and fainter sources that are present in the sky.  
Values of the cumulative counts for different S/N levels are tabulated in Table \ref{tab:intcounts}.  This gives an appreciation for the number of sources that could be detected at various significance; these are of the order of hundreds of thousandths for 560 and 1400\,MHz and of hundreds at 9200\,MHz for the 1000\,h images.

\begin{table}
   \caption{Number of sources in the simulated maps above different noise levels.  \label{tab:intcounts}}
    \centering
    \begin{tabular}{c|c|c|c|c}
Exposure&$\sigma$ levels&560\,MHz&1400\,MHz&9200 MHz\\
\hline
1000h&5&757985&227168&657\\
&7&624389&155426&295\\
&10&410832&126442&208\\
\hline
100h&5&257472&80436&123\\
&7&199833&47636&87\\
&10&114085&25849&74\\
\hline
8h&5&60472&12763&45\\
&7&29422&8698&40\\
&10&20156&3934&29\\
\end{tabular}
\end{table}

\subsection{The challenge}
The SKA community was invited to retrieve the SDC1 images and undertake source finding, source classification and characterization. The results submitted should be catalogues of detected sources, containing:

\begin{enumerate}
\item	Source coordinates (RA, Dec) to locate the centroids and where appropriate the core positions;
\item	integrated primary-beam corrected flux density;
\item	core fraction (it is different from zero only for AGN);
\item	major and minor axis size;
\item	major axis position angle;  
\item	Source population identification (one of AGN-steep, AGN-flat, SFG)
\end{enumerate}

Potentially challenging aspects of this dataset are:
\begin{itemize}
\item The sheer number of sources (see Table \ref{tab:intcounts}), which requires the source extraction and source characterization methods to be automated, efficient and, ideally, parallel.
\item The source density, which allows probing source extraction in a regime of high source crowding.
\item Within each SDC1 map, and even more so across the 950--9200\,MHz frequency range, sources range from unresolved to fully resolved, revealing in some case a complex morphology. Such diversity could challenge source extraction methods. 
\end{itemize}
The SKA imaging products will be challenging in terms of data sizes as well. For this exercise, we considered only one frequency per band, thus reducing the size to $\sim$4GB for each map. Although not representative of the full SKA data size, this file size already requires participants to access and analyse portions of the data separately on most computers, and therefore to organise the workflows in a way that can be scaled to even bigger sizes.

The description of how submission were evaluated and scored in given is Sec. \ref{sec:evaluation}

\section{Methods}\label{sec:methods}
The SDC1 dataset was released and advertised publicly on 25th November 2018. Several teams, from various countries in the world, registered their interest on analysing the dataset. Of those, 9 teams managed to submit results by the deadline of 30$^{\rm th}$ April 2019, and 8 participated to this paper.
Participation in the challenge was completely open and voluntary. Different teams approached it from a different level of specific expertise and preparedness. Teams were allowed to use their own developed methods as well as publicly available ones. As a result, the list of approaches used is most likely not complete.
This section lists the teams in alphabetical order and describes their analyses. 

\subsection{ARCIt-CACAO (Burkutean, Brand, Massardi, Schisano, Bonato, Liuzzo,
Marchili, Giannetti, Rygl)} \label{sec:cacao}

ARCIt-CACAO is a source detection and classification pipeline developed in the context of the first SKA data challenge \footnote{\url{https://www.ira.inaf.it/cacao/CACAO\_V1/cacao.html}}. It is entirely Python based, parallelized and makes use of SciPy \citep{2020SciPy-NMeth} and Astropy \citep{astropy:2013, astropy:2018} routines. The ARCIt-CACAO pipeline is split into three distinctive steps, namely source detection, description and classification.

For the source detection and description steps, ARCIt-CACAO uses identical pipeline set-ups for all bands and exposure times, the only differences being the band, exposure time and associated expected theoretical noise, and of course the cleaned images and their primary beams. All other parameter settings were
held constant for each ARCIt-CACAO run with a particular combination of
frequency and exposure time. This decision was made in order to mimic
a general solution approach to the problem rather than a solution
tailored to a band-exposure-time combination.

The cleaned input images were split into sub-fields of 1024$\times$1024\,pixel\textsuperscript{2} so that the source detection and description could be run in parallel on each of them. 
Although producing a different set of sub-images that overlap with the
originally generated sub-images can be implemented in the ARCIt-CACAO
pipeline via an additional tessellation, this was not done for the SKA
challenge in order to speed up the overall potential source table
generation per run. As the ARCIt-CACAO source detection operates at the
sub-image level, this could have resulted in some extended high-flux
sources, which have a higher probability of spanning different
sub-images, not having been recognised.
Solutions from the outer 512 pixels on the sides of the large 32768$\times$32768\,pixel\textsuperscript{2} input images were rejected to avoid edge effects.

The ARCIt-CACAO pipeline firstly applies a primary beam correction to
the cleaned input sub-fields, using a re-sampled primary beam that
matches the spatial extent and resolution of the cleaned sub-images. A gaussian filter with a kernel size set to the FWHM of the clean beam is applied to the sub-field input images leading to the creation of a binary mask after filtering out values smaller than 0.5\,$\sigma$, where $\sigma$ corresponds to the expected theoretical noise level input parameter. Islet identification within the binary mask via SciPy tasks \citep{2020SciPy-NMeth} generates a list of potential source locations and maximum 2D pixel source extent for each potential source within each sub-field.

For each of these islets within the sub-field, the ARCIt-CACAO source description routine generates source description measures from image moment analyses of the sub-field region spatially extracted at the islet region. The position angle and source extent were computed from the eigenvectors and associated eigenvalues of the covariance matrix of the islet region extracted from the primary beam corrected sub-field and the flux was determined in the selected region. Potential sources with a maximum extent smaller than the beam size were rejected. In addition, a conservative criterion of source acceptance was placed via a 2\,$\sigma_{\rm pb}$ threshold on the maximum pixel value within each islet, where $\sigma_{\rm pb}$ corresponds to $\sigma$ divided by the mean primary beam correction factor within the islet. 

For each sub-image, individual detected source lists were created that were then concatenated into the final table containing source position, size and flux using pandas \citep{mckinney-proc-scipy-2010}. Source classification was made based solely on the source flux and size distribution property analyses of the originally provided training data sets.

\subsection{Engage SKA -- Portugal}
Using the function \texttt{read.fits} from the \textsf{R} package ``astro'', we divided the simulated SKA continuum images into smaller ones (of 16384 $\times$ 16384 pixels). The primary beam images were re-projected and re-sampled  in order to match the dimensions of the latter continuum sub-images, using the function \texttt{reproject\_interp} of the package ``reproject'' from Astropy \citep{Astropy2013, Astropy2018}. Then, using the CASA \citep{McMullin2007} task \texttt{impbcor}, we obtained the primary beam corrected images.

We ran \textsc{ProFound} \citep{Robotham2018, Hale2019} on the the primary beam corrected images to pinpoint the sources and determine their properties. The argument \texttt{skycut}, defining the S/N threshold, was set to 3, which is the minimum value typically used in extragalactic astronomy. 
The minimum of pixels required to identify a source was chosen to be 2 (\texttt{pixcut}$=2$), which is the minimum value that allows an orientation of the source to be defined. Finally, the argument \texttt{tolerance} was set to 1 in order to maximize the deblending of nearby sources. 

With the Tool for OPerations on Catalogues and Tables (TOPCAT) software \citep{2005ASPC..347...29T} we searched for the counterparts of the identified sources, via a cross-match of the simulated multi-frequency data. Based on the latter  information we classified the sources as either  steep-spectrum AGN (those objects that have a spectral index $\log (F_{1400}/F_{9200})$ larger than 0.7); flat-spectrum AGN (having spectral index $<0.7$); and as star-forming galaxies (if not detected at 9200\,MHz).

\subsection{hs (Lukic, Br\"uggen, De Gasperin)}

We explored source finding using ConvoSource, a Convolutional Neural Network (CNN) that was trained on a solution map derived from knowledge of the source locations. For the purpose of source finding, the output images that had to be matched are those of the source locations, rather than the original input source maps. For that purpose, we transformed the source locations into an image, the source location map. This map, as well as the original source map, was segmented into smaller square images of 50 $\times$ 50 pixels, which then provide the inputs to the CNN.

Our work was based on the {\tt Keras} package  with the {\tt TensorFlow} backend. We used a convolutional network architecture of three consecutive convolutional layers and one dense layer. Altogether, this network has a total of 32,193 parameters.
Early stopping was used with a patience of five training epochs. Eighty percent of the data was used for training, and the remaining 20 percent was used for testing. In order to make the network more robust by reducing overfitting, we placed  dropout layer with a dropout fraction of 0.25 between the first and second convolutional layer. The batch size was set to 128 and we also used the binary cross-entropy cost function. 
We also experimented with augmentation by rotating and flipping images.
More details of the algorithm can be found in \cite{2019Galax...8....3L}.

\subsection{ICRAR (Wu, Wong \& Dodson)} 
The ICRAR team's method is
primarily based on the Classifying Radio galaxies
Automatically using Neural networks \citep[ClaRAN; ][]{wu19} version 0.2\footnote{\url{https://github.com/chenwuperth/ClaRAN}}
prototype.  In this subsection, we provide a brief description of ClaRAN
and our specific approach to SDC1.

ClaRAN's primary purpose is to classify extended radio sources within
any given image field without knowing a priori the number of independent
radio sources within that image --- thereby combining the two tasks of source
identification with source morphology classification. In recent years, citizen
science projects such as Radio Galaxy Zoo \citep[RGZ; ][]{banfield15} has helped
increase the sample sizes of extreme classes of extended radio galaxies \citep[e.g.\ ][]{banfield16,kapinska17}.  The many RGZ-enabled discoveries and
publications using archival radio observations from the FIRST survey \citep{becker95} in combination with WISE mid-infrared maps \citep{wright10} demonstrate that current methods for radio source classification and our understanding of extended radio galaxies can still be furthered.  To this end, we developed ClaRAN, a prototype end-to-end deep learning classifier, trained on extended radio sources identified in a 2018 version of the RGZ Data Release 1 catalogue \citep{wong20}.  ClaRAN employs a deep learning method based on a Faster R-CNN algorithm \citep{ren15} that has been reimplemented in {\tt TensorFlow}. 
A more complete description of ClaRAN can be found in \citet{wu19}.

The main update between the method as described in \citet{wu19} (version 0.1) and the one used in the challenge (version 0.2) is that the default network architecture is now based on the ResNet50 \citep{he15} model, instead on the VGG-16 \citep[configuration D; ][]{simonyan15}.  
 Version 0.2 has integrated Feature Pyramid Networks for object detection \citep{lin16} which enables multi-scale feature extraction and also supports on-the-fly rotation within the extended image augmentation pipeline.  These new updates have attained an improved mean average precision (mAP) from 82.6~per-cent with version 0.1 to 86.1~per-cent with version 0.2 for ClaRAN's D3 method (which involves the input of an infrared map overlaid with radio contours).

Specific to SDC1, our approach can be generalised into three main steps:
1) the pre-processing and preparation of the training datasets; 2) training
ClaRAN; 3) source and flux extraction.

Before we begin the training, we need to prepare and pre-process the ground truth catalogues provided by SDC1 in order to obtain a suitable training set for  ClaRAN.  The pre-processing steps are as follows:
\begin{itemize}
  \item Convert the catalogues to CSV format files;
  \item Divide images into a set of smaller cutouts that are 205 pixels by 205 pixels in size;
  \item Filter for sources from the ground truth catalogues that have fluxes ($S$) greater than $k \times \sigma$, where $\sigma$ is the respective RMS of the image field) and  $k$ ranged from 0.5 to 3.0;
  \item Determine the bounding box and class label for each filtered source and put into a JSON file with the source and cutout identifiers.
\end{itemize}

The pre-processed dataset is divided into two: a JSON file for the training set and another one for the testing dataset.  Using ClaRAN version 0.2, we train ClaRAN to recognise the types of sources that are presented in SDC1.  

After running ClaRAN on the full SDC1 dataset, we measure the fluxes of the identified sources  automatically using the {\tt Imfit} tool from the {\sc{MIRIAD}} software suite \citep{sault95}. As ClaRAN was built specifically for source classification, we needed to estimate the integrated fluxes of the sources with a different tool. While we initially fitted a Gaussian to each radio source identified by ClaRAN, we found a systematic underestimation of the true flux and thus we integrated the flux of the radio sources within a flat-top elliptical disk, which had significantly better performance.
All materials used by our team's solution for SDC1 is in a publicly-accessible Github repository at {\url{https://github.com/ICRAR/skasdc1}}.  

\subsection{IPM (Goodarzi, Bagheri, Sabz Ali, Saremi, Sheikhnezami, Vafaei Sadr, Zhoolideh Haghighi, Wong)}
At IPM, we follow two different paths to find radio sources in the image. The first approach, which we labelled IPM1, used a pipeline called SExtractor \citep{1996A&AS..117..393B}. The second approach, labelled IPM2, is a method we have developed from scratch by using image processing tools such as scikit-image \citep{van2014scikit}. 

 SExtractor is a free pipeline to extract catalogues from astronomical images. Although this routine was originally developed to identify objects from large-scale galaxy-survey data, it can also extract sources from crowded star fields \citep{1996A&AS..117..393B}. In order to use this package, some parameters such as detection threshold, detection minimum area, should be set. We selected these parameters by trial and error and visual inspection. First, we start with the guess values and run the software. Then, the calculated contours were plotted and overlapped with the image. We iterated on the input parameters until the number of detected objects and recognized shapes reached an optimal level, as judged by visual inspection. By using this method, we set the detection threshold at 2.5\,$\sigma$ above the local background and choose 3 pixels as the minimum number of pixels above threshold value in which a group should have to be detected as a single object. 

The SExtractor software utilizes multi-thresholding as a de-blending method, so we set the number of de-blending sub-thresholds as 32 (the value recommended in the software manual).  Also we selected the minimum contrast parameter for de-blending as 0.005. 
Since the simulated images were crowded, we selected the option of local background in order to compute background in a rectangular region around the isophotal limits of the object. This helped to improve the evaluation of the background and the consequent photometry.
We should note that our method is the initial approximation for usage of the SExtractor as source finder; by comparing its results with the true catalogue, it is possible to optimize the input parameters and improve the results. The results of this technique are presented in section \ref{sec:results} as IPM1.

\begin{figure}
\centering
\includegraphics[width=0.5\textwidth]{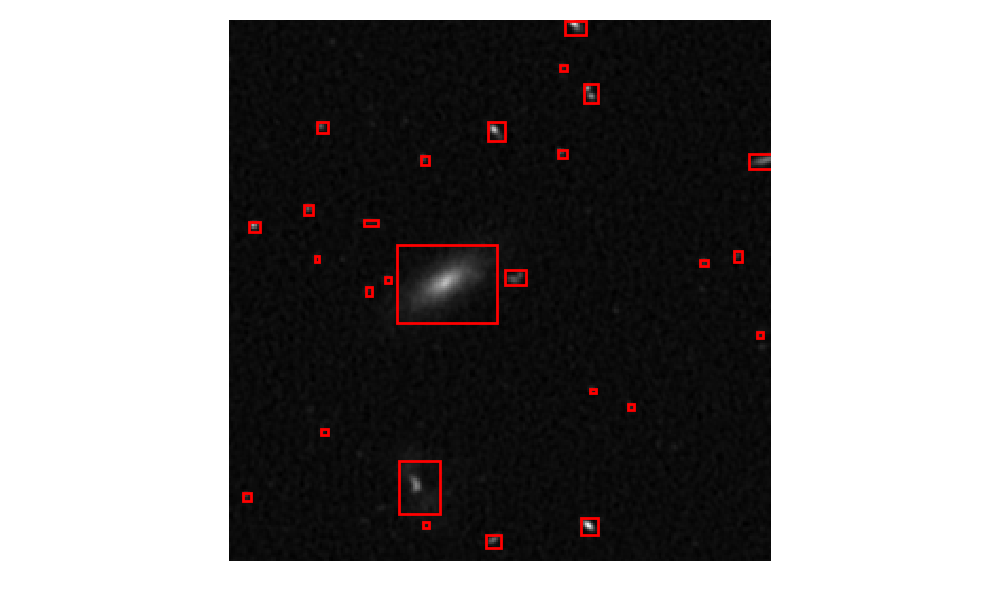}
\caption{Detected sources with their bounding boxes in a 256 by 256 pixel part of the original image.}
\label{fig:bounding_box}
\end{figure}

In the second approach, we started with a denoising step to reduce the noise levels in the images. Amongst the many denoising methods available, we used the median filter module available in Scipy \citep{SciPy}, and by applying a multi-dimensional median filter on the original images, we removed the background noise. Then, in order to distinguish the sources from the background we have to set a threshold. The best thresholding method for our images, as assessed by trial and error,  resulted to be Li thresholding \citep{LI1998771}. After denoising, we make a binary (black and white) image in which radio sources are white objects and background would be black. Finally, we use some modules of the scikit-image \citep{van2014scikit} including {\tt morphology} and {\tt measure.label}, {\tt measure.regionprops} to find the centroids of radio sources alongside their bounding boxes. Once we get bounding boxes, not only we can calculate the properties required by the challenge, but also other interesting features such as Euler number and eccentricity. The results of this technique are presented in section \ref{sec:results} as IPM2, and for a better visualization of the bounding boxes and recovered sources you can refer to Fig \ref{fig:bounding_box}. For future prospects, we investigate training convolutional neural networks as proposed in \citep{vafaei2019deepsource} as denoising method in order to customize the process. 

\subsection{JLRAT (Yu, Liu)}

The size and intensity of the radio sources provided by SDC1 both have a large dynamic range. To deal with these features and detect weak sources, we have designed the JLRAT source finder model (JSFM) with feature pyramid network. 
Our method can be described in four main steps: (1) pre-processing, (2) denoising model, (3) source detection and classification, and (4) source property characterization. 

\begin{figure}
\centering
\includegraphics[width=0.35\textwidth]{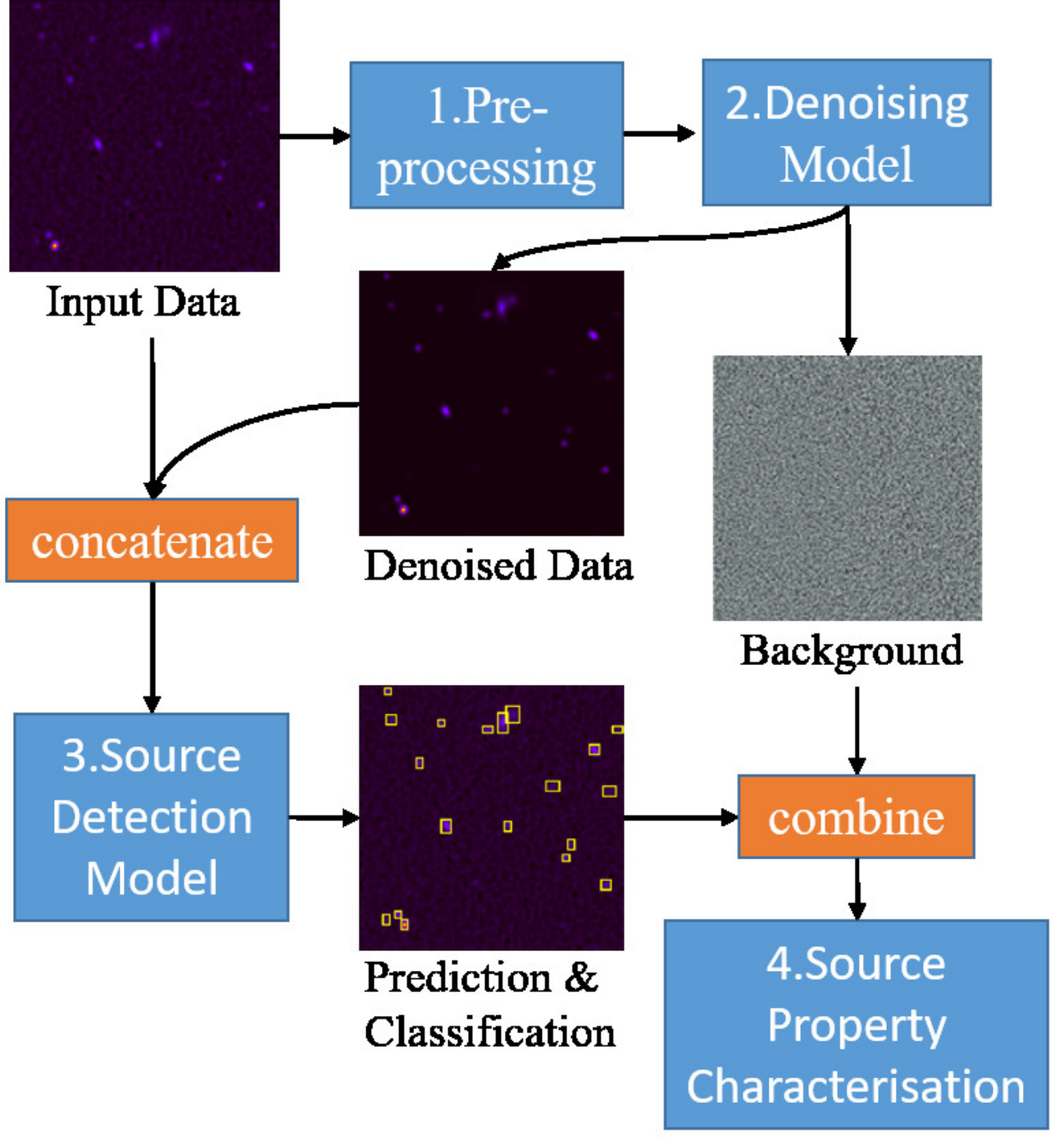}
\caption{The pipeline of JLRAT source finder model}
\label{fig:pipeJLRAT}
\end{figure}

The pipeline of our method is demonstrated by Fig. \ref{fig:pipeJLRAT}. In the first place, the raw image is pre-processed with pixel value scaling, the $\log_{10}$ operation and zero centering. Then, we build a denoising model, which applies the core idea from \cite{zhang2017beyond}, but without batch normalization \citep{Ioffe2015BatchNA} in convolutional layer.
By using the residual learning, it generates both smoothed and background image.
The RMS of the background data times a scalar is the threshold value for building a binary image that is used to source property characterization. In the third step, the source detection model locates sources in the image and identifies their corresponding classes. 
Finally, the binary image of each detected source region is fit to extract the source properties.

The model for the noisy observation $S_{\rm obs}$ is $S_{\rm obs} =S_{\rm real} + n$, where $S_{\rm real}$ is a clean image and the noise $n$ is additive white Gaussian noise (AWGN) with standard deviation $\sigma$. The purpose of image denoising is to recover a clean image from the noisy observation $S_{\rm obs}$. In this section, we use a multi  convolutional layer as our denoising model that adopts a residual learning to train a residual mapping $\mathcal R(S_{\rm obs}) \approx n$. That acquires a background image and generates a denoised image $S_{\rm real}=S_{\rm obs}-\mathcal R(S_{\rm obs})$. Formally, the loss function is:
$$
\ell(\Theta)=\frac{1}{2 N} \sum_{i=1}^{N}\left\|\mathcal{R}\left(S_{\rm obs}^{i} ; \Theta\right)-\left(S_{\rm obs}^{i}-S_{\rm real}^{i}\right)\right\|_{F}^{2}
$$

A key idea of our method of finding source location is that we evaluate it on different scales. 
This kind of method has been applied in various scenarios, such as \cite{Cai2016AUM}, \cite{articleFPN}, \cite{Redmon2018YOLOv3AI},  \cite{zhao2019m2det} and \cite{inproceedingsMSSM}. In order to retain the features of relatively small sources, our model does not use many down-sampling operators in the base network. 

The source detection model begins with a simple Base Network to extract the base feature. Then, a Feature Pyramid Network is applied to build a multi-scale feature pyramid, with two alternative models. %which is generated by two models with alternatively used. 
The maximum size layer of each feature pyramid is the output who will concatenate with the base feature as the input for the next stage. After that, feature channels concatenate and weight model (FCCWM) firstly concatenate same scale layers among all feature pyramids, then weight each scale channels separately through the squeeze-and-excitation method from \cite{articleSE}. 
The overall architecture of source detection model and feature pyramid model I and II are illustrated in Fig. \ref{fig:m1m2JLRAT}.

In the prediction step, in order to achieve source location regression and classification, six different scale features are connected with five convolution layers respectively. The detection boxes scale ranges for those features, we follow the setting of the original Single Shot MultiBox Detector from \cite{liu2016ssd}.

After the location of the sources is acquired, the RMS of the background which is generated by the Denoising Model is used as a benchmark of the threshold value. Then it is multiplied by a scalar to determine the binary image that is used for the source Property Characterisation.
\subsection{RADGK (Pritpal, Pankaj, Mohit, Prabhakar)}
The RADGK team approached the challenge without having an automated data analysis pipeline. At this stage we concentrated on understanding the single steps involved in the analysis, by working on the few sources that could be identified by visual inspection and interactive processing. Further steps including automation of our procedures are outlined at the end of this section. 

The simulated dataset was downloaded at 560 and 1400\,MHz and for 1000 hours exposure. 
Sources were identified by visual inspection and the license-based ALADIN \citep{2000A&AS..143...33B} software has been used to extract their properties, in particular position (core and centroid), peak and integrated flux, major and minor axis and position angle.  We identified 34 sources at 560\,MHz and 25 sources at 1400\,MHz by visual inspection.

The centroid of each source was calculated using the following expression:
\begin{equation}
    \big(x_o,y_o)=\Big(\frac{\Sigma I_{ij}x_{ij}}{\Sigma I_{ij}},\frac{\Sigma I_{ij}y_{ij}}{\Sigma I_{ij}}\Big)
\end{equation}
where $x_{ij}$ and $y_{ij}$ indicate the RA and Dec coordinates of the pixel $(i,j)$ and $I_{ij}$ is the flux density at that location. The list of pixels $i,j$ to consider for each source was selected by cropping the individual sources from the image file in FITS format using the ALADIN® software. Data extraction for each source is obtained in tabulated form again using ALADIN®, where the first column gives the intensity/pixel and the second and third columns give the position of the pixel in RA and DEC.

We next fitted a 2D elliptical Gaussian function centred at $(x_o,y_o)$ for each detected source, given by
\begin{equation}
    I(x,y)=A_{o}e^{\Bigg\{-\Bigg[\frac{\big(x-x_o\big)^2}{2\sigma_x^2}+\frac{\big(y-y_o\big)^2}{2\sigma_y^2}+\frac{\beta\big(x-x_o\big)\big(y-y_o\big)}{\sigma_x \sigma_y}\Bigg]\Bigg\}}.
\end{equation}
 Here $I(x,y)$ is the intensity at coordinates (x,y), $A_o$ is the amplitude (maximum intensity), $\sigma_x$ and $\sigma_y$ are the standard deviations along the major and minor axis respectively  and $\beta$ is the source position angle. All the four parameters $(A_o, \sigma_x,\sigma_y, \beta)$ have been obtained by fitting the Gaussian function to the source intensity profile. The beam major axis (BMAJ) and beam minor axis (BMIN) have been obtained as the FWHM along the major and minor axis, derived from the Gaussian model. In order to obtain the integrated flux we first correct the primary beam data using CASA task {\tt impbcor} and we then integrate the primary-beam corrected flux over all the pixels associated to the source.

As explained above, so far we have extracted sources only through visual inspection. An automated extraction, which may be applied in future, would consist of the following steps:
\begin{itemize}
    \item[1.] The original image will be split into several smaller regions.
    \item[2.] The software PyBDSF \citep{2015ascl.soft02007M} will be used to generate the catalogue of each
sub-region.
\item[3.] After source extraction from PyBDSF, we will identify different types of sources, such as,  SS-AGN, FS-AGN and SFG.  The visual extraction will be used to validate the automated extraction procedure.
\item[4.] Our current pipeline will be used to extract different source properties.
\end{itemize}
For extended sources we may also use the Gaussian Mixture model (GMM) to extract the source information with better accuracy.

\subsection{Shanghai (An, Jaiswal, Lu, Mohan, Lao)}
The simulated images were analyzed through three source-finding algorithms, namely AEGEAN \citep{hancock2012,2018PASA...35...11H}, DUCHAMP \citep{2012MNRAS.421.3242W} and SExtractor \citep{1996A&AS..117..393B}. These methods extract contiguous islands of pixels above a specified detection threshold in the image. After applying a suitable fitting algorithm (e.g. two-dimensional Gaussian models) these islands are referred to as astronomical objects. Normally, the detection threshold is expressed in multiples of RMS noise in the image. Each software provides the centroid positions of the extracted sources, their integrated flux densities and their geometrical properties. The integrated flux densities were finally corrected for primary beam by dividing with the primary beam response at each source coordinates. 

The number of sources extracted through DUCHAMP was found to be significantly less in comparison to the other two for almost the same detection threshold level used (5\,$\sigma$), therefore the source finding and comparison were performed through AEGEAN and SExtractor only. Extracted sources from AEGEAN and SExtractor were first cross-matched with the training data set using TOPCAT \citep{2005ASPC..347...29T}. The extracted source positions from SExtractor were found to match with the actual source positions more accurately than that from AEGEAN, so we finally used the results obtained with SExtractor for the final submission. The integrated flux densities estimated above were corrected for the primary beam by dividing with primary beam response (the given ancillary data) at the centre of the source. The source geometrical parameters, {\it i.e.}, major axis length, minor axis length and major axis position angle were estimated by fitting the source brightness distribution with a 2-Dimensional Gaussian model having elliptical footprints. We have not attempted to fit different models for different sources according to the various source classes (SS-AGN, FS-AGN, or SFG).

\section{Evaluation of the submissions} \label{sec:evaluation}
\begin{figure*}
\includegraphics[width=7.5cm]{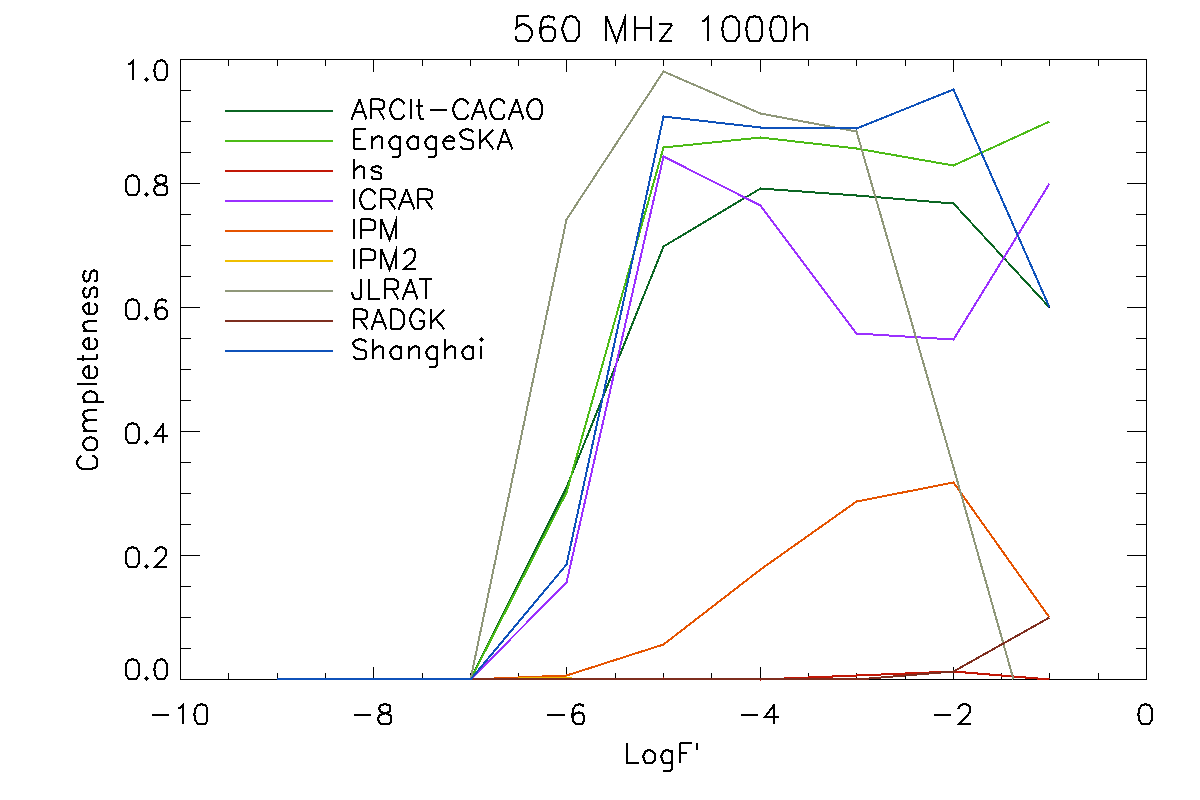}
\includegraphics[width=7.5cm]{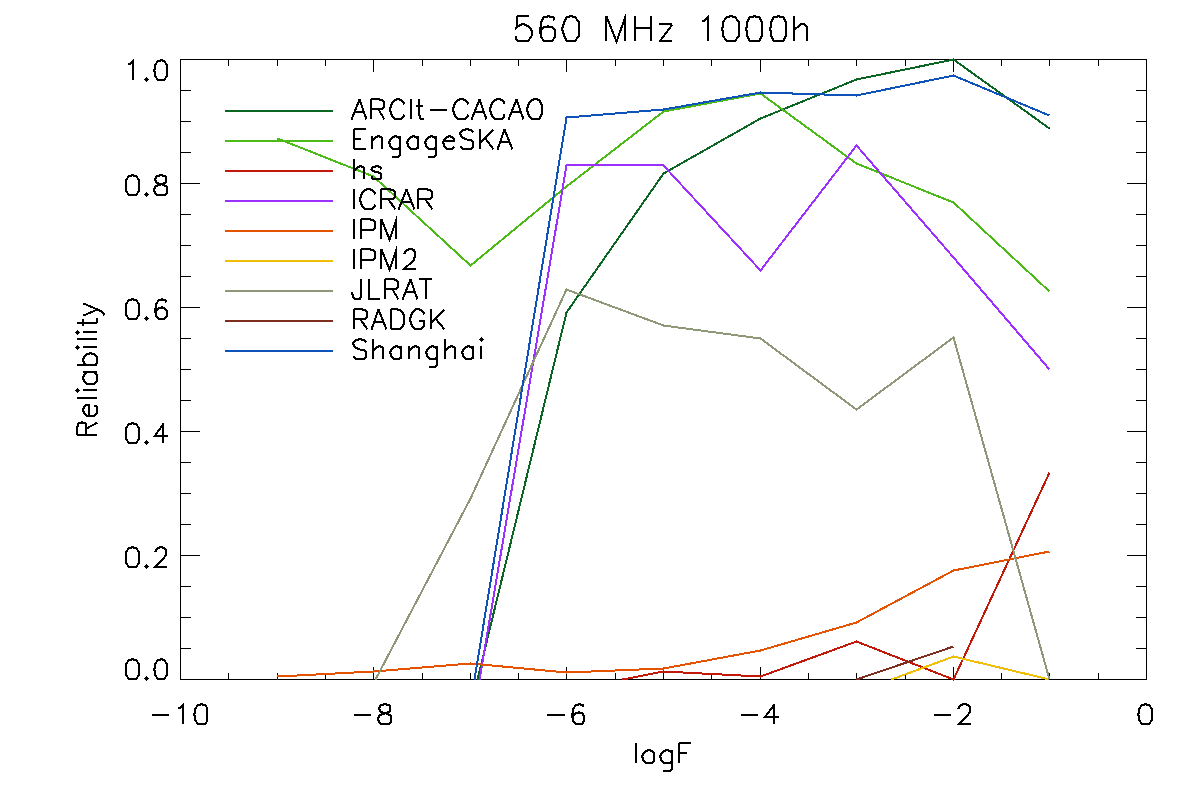}

\includegraphics[width=7.5cm]{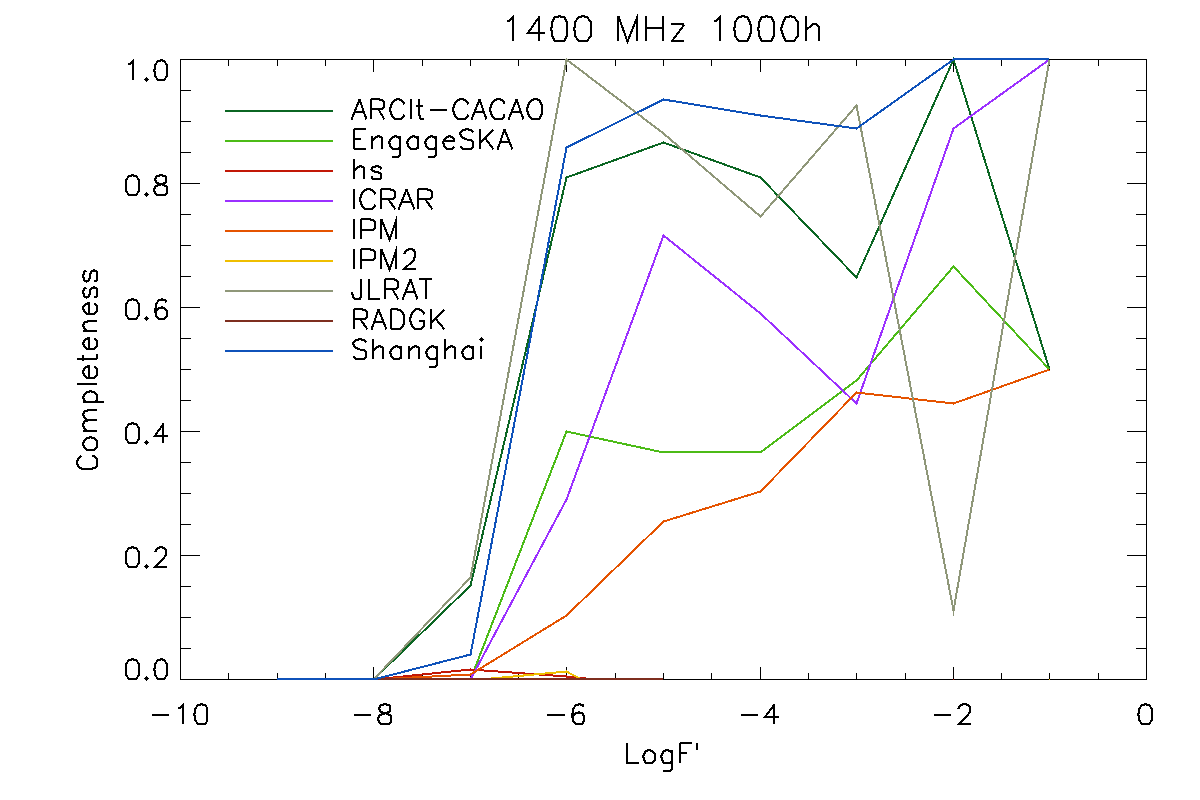}
\includegraphics[width=7.5cm]{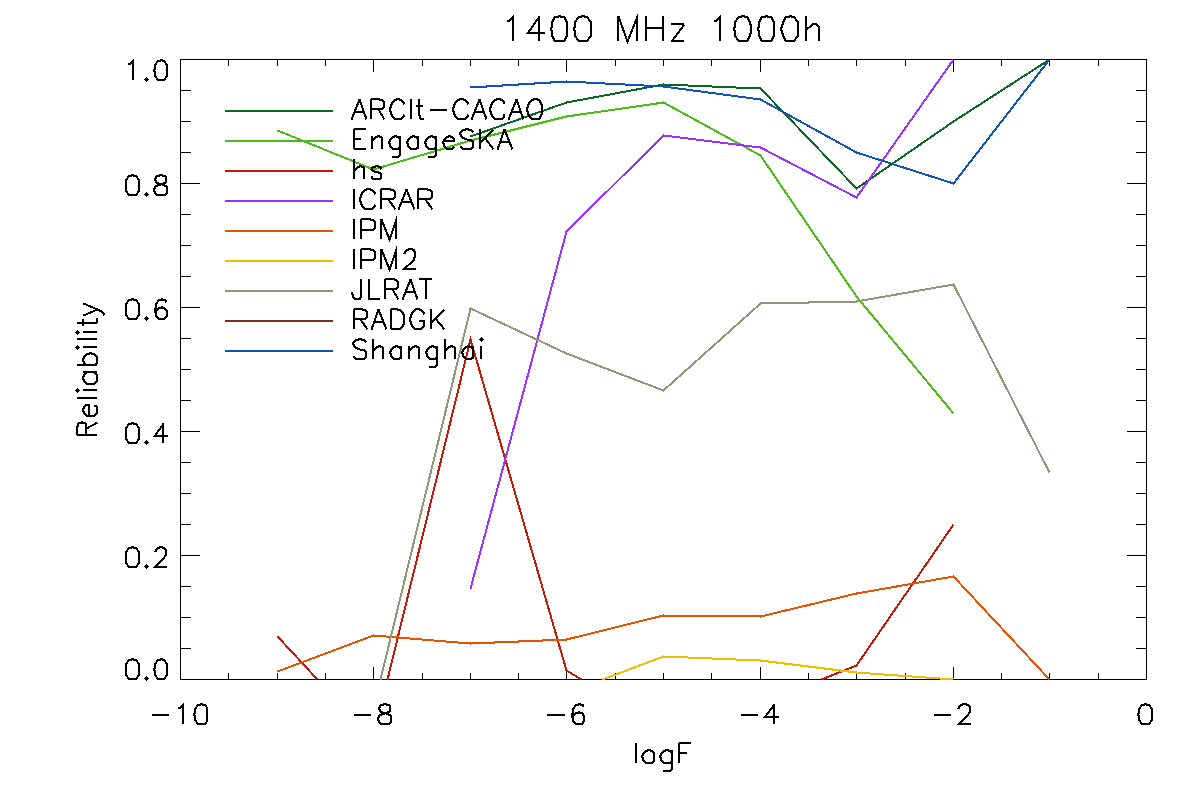}

\includegraphics[width=7.5cm]{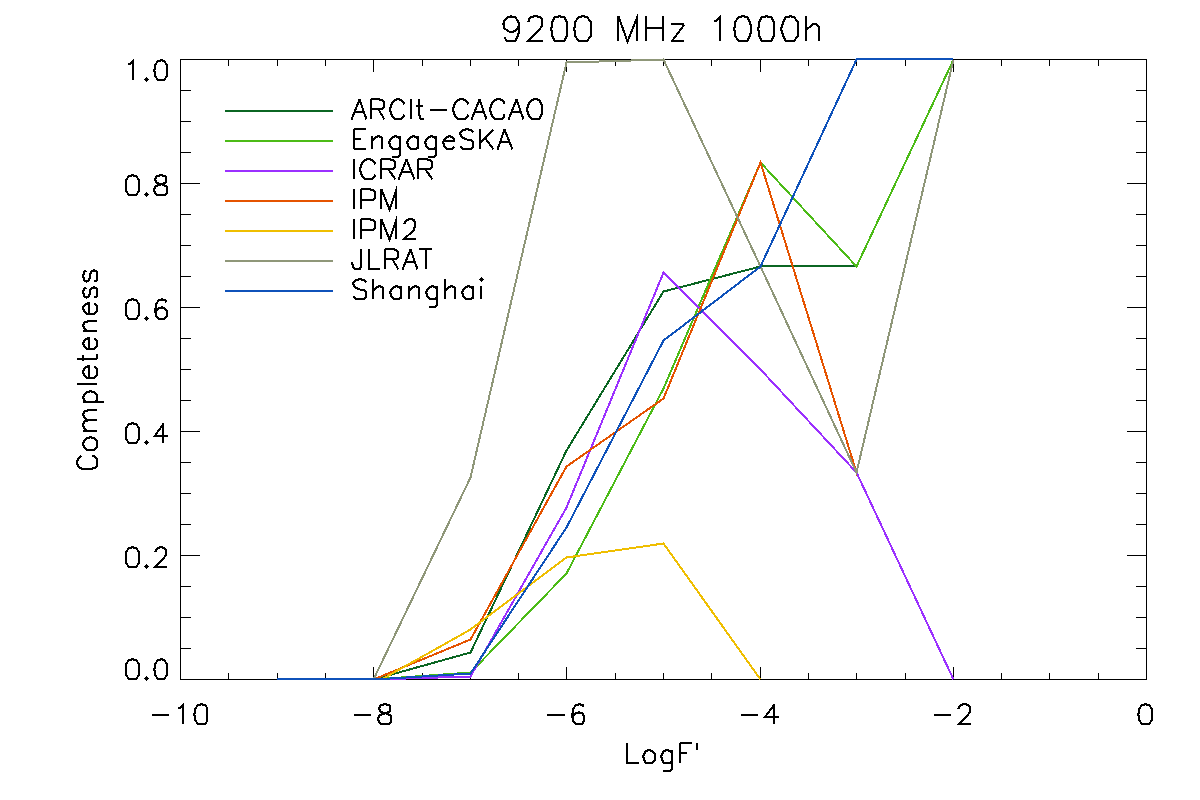}
\includegraphics[width=7.5cm]{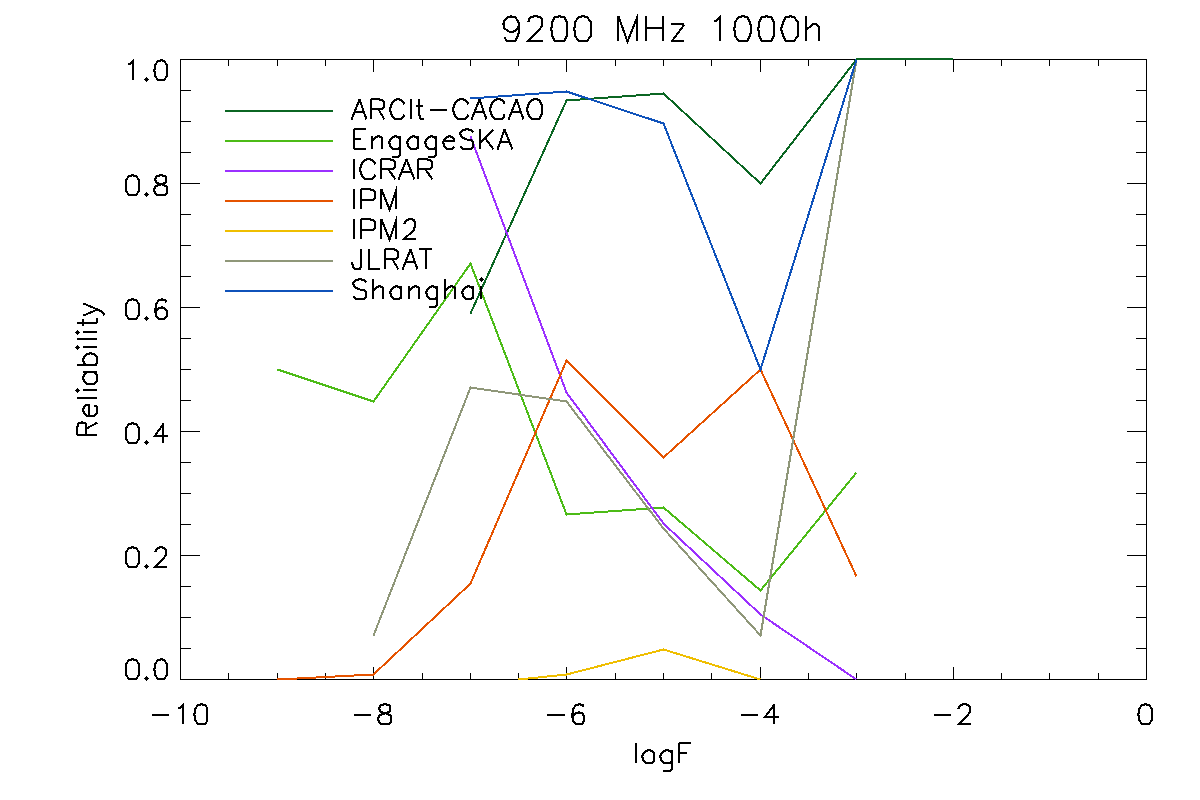}
\caption{Completeness (left) and reliability (right) as a function of frequency (560\,MHz, 1.4\,GHz and 9.2\,GHz from top to bottom) for all the 1000\,h submission sets.}
\label{fig:CandR_teams}
\end{figure*}

\subsection{True and Submitted catalogue crossmatch} \label{sec:cross}
The first step of the evaluation analysis was to crossmatch the True and the Submitted catalogues. 
We started by performing a positional cross-match to identify true sources within a given radius from each detection (we used 1.5 the estimated source size convolved by the beam).

Given the very high density of sources in the True catalogue (50 sources per square arcmin at 560\,MHz on average) this step typically yields multiple matches. The best match among this list has been defined as the source minimizing the sum in quadrature of positional, size and flux density mis-match, respectively defined as

\begin{eqnarray}
D&=&\sqrt{D_{\rm pos}^2+D_{\rm size}^2+D_{\rm flux}^2}\label{d}\\
D_{\rm pos} &=& \sqrt{(x-x')^2+(y-y')^2}/S'  \label{dpos}\\
D_{\rm size} &=& |S-S'|/S'\label{dsize}\\
D_{\rm flux} &=& |f-f'|/f'  \label{dflux}
\end{eqnarray}
where the prime denotes the attributes of the True catalogue; ($x,y$) are the pixel coordinates corresponding to (RA,DEC); $S$ is the average source size, computed as the mean between major and minor axis, and $f$ is the integrated flux density. We found this method to be much more reliable in identifying the best match than using position only, given the field crowding. 

In setting the threshold on $D$ to finally accept the best match as a true positive, we had to consider the probability of a false positive to be associated to a true source by chance. We assessed this likelihood by performing a null test for each submission. This consisted of creating a copy of the Submitted catalogue with random coordinates (which we called Null-test catalogue) and repeating the cross-match. The matches yielded by this procedure are all chance matches, and provide an estimate of the contamination due to chance matches for that submission. 
We found that, due to the high number of true sources in the catalogue, chance match could be significant. We set the threshold on $D$ to 5\,$\sigma$ of the distribution across all teams, after which this contamination is below 10\% for most submissions (see Table \ref{tab:detections}).

\subsection{Scoring metrics}\label{sec:scores}
SDC1 addresses two separate aspects: source finding and source classification/characterization. In the following we define metrics to judge these aspects individually, as well as a global score that considers them both. Each metric distils the performance to a single number per submission set (three frequencies, one depth).

To evaluate the source finding component of SDC1, we defined scores based on the number of detections $N_{\rm d}$ and the number of matches $N_{\rm m}$
\begin{eqnarray}
C_{\rm tot}&=&\frac{N_{\rm m,\nu1}}{{\rm FoV}_{\nu1}}+\frac{N_{\rm m,\nu2}}{{\rm FoV}_{\nu2}}+\frac{N_{\rm m,\nu3}}{{\rm FoV}_{\nu3}} \label{ctot}\\
R_{\rm tot}&=&\frac{1}{3}\left[ \frac{N_{\rm m,\nu1}}{N_{\rm d,\nu1}}+\frac{N_{\rm m,\nu2}}{N_{\rm d,\nu2}}+\frac{N_{\rm m,\nu3}}{N_{\rm d,\nu3}}\right],
\end{eqnarray}

where $\nu1$, $\nu2$, $\nu3$ are the three frequency channels for the same depth and FoV$_{\nu}$ is the sky area at frequency $\nu$. The field of view normalizations in eq. (\ref{ctot}) are necessary to give comparable weight to the three frequency channels despite the very different number of sources in them. $R_{\rm tot}$ is the total reliability averaged between the three frequencies, and as such it is already correctly normalised.  

To assess the accuracy of source classification and characterization, we consider errors on all the measured source attributes. These include 
\begin{eqnarray}
D_{\rm cf} &=&|\rm{cf}-\rm{cf'}|/0.75  \label{dcf}\\
D_{\rm PA} &=& |PA-PA'|/10^\circ \label{dpa}\\
D_{\rm class} &=& 0 \rm{\;if\;correctly\; classified}\label{dclass}\\
&=&1 \rm{\;otherwise}\nonumber
\end{eqnarray}
where PA is position angle, cf is core fraction and the prime is the True catalogue value,
as well as the already defined $D_{\rm pos}$, $D_{\rm size}$ and $D_{\rm flux}$ [eqns (\ref{dpos})--(\ref{dflux})].

Those errors were used to associate a weight per matched source, $w_i$, ranging from a minimum of 0 to a maximum of 1 (see Appendix \ref{sec:wi} for more details). Once summed over all matched sources, this yields an "effective" number of matched sources,
\begin{equation}
    \tilde N_{\rm m}=\sum_{i=1}^{N\rm m} w_i \le N_{\rm m},
\end{equation}
which is weighted down by errors in their characterization/classification. We finally defined an accuracy metric as
\begin{equation}
A_{\rm tot}=\frac{\tilde N_{\rm m,\nu1}}{{\rm FoV}_{\nu1}}+\frac{\tilde N_{\rm m,\nu2}}{{\rm FoV}_{\nu2}}+\frac{\tilde N_{\rm m,\nu3}}{{\rm FoV}_{\nu3}}.
\end{equation}
 
The global SDC1 score, to assess both accuracy and source finding, is finally defined as: 
\begin{equation}
G_{\rm tot} = \frac{B_{\nu1}}{{\rm FoV}_{\nu1}}+\frac{B_{\nu2}}{{\rm FoV}_{\nu2}}+\frac{B_{\nu3}}{{\rm FoV}_{\nu3}} \label{gtot}
\end{equation}
where $B$ is the difference between the effective number of matches and the number of false positives, $N_{\rm f}$:
\begin{equation}
    B=\tilde N_{\rm m}-N_{\rm f}.
\end{equation}
We note that $B$ (and therefore $G_{\rm tot})$ can become negative, if the number of false detections is larger than the number of matched sources, or if the accuracy of characterization/classification is low. 

It is worth pointing out that, due to the way that our cross-matching procedure works, the source finding and the source characterization performances are linked. Sources that have been correctly identified by a source finding pipeline could fail the cross-matching step due to a large error on position, flux or size. In this case, they would be classified as false positives, with an impact on the source finding metrics as well as the accuracy ones. Although this feature of the evaluation process is not ideal, it could not be avoided. Setting a generous allowance for errors in the cross-match results in a very high contamination from chance matches, which makes every metric meaningless. 

In the case of partial submissions (some of the frequency catalogues are missing for a given depth) a score of 0 was awarded for the missing frequency component in all scores.

The Leaderboard\footnote{The Leaderboard is available at \url{https://astronomers.skatelescope.org/ska-science-data-challenge-1-results/}} for SDC1 was based on $G_{\rm tot}$ achieved by each team on the 1000\,h submission by the challenge deadline, 30$^{\rm th}$ April 2019. Since the deadline, teams have been allowed to update their submission for the purpose of this paper.

One of the teams that participated in the challenge (Ox-ICRAR) made the choice not to have their results appear in this paper, however they are listed in the original Leaderboard. 
The results presented in the next section consider all the metrics defined in section \ref{sec:scores} and additional diagnostic plots. 

The scoring procedure described in this section was originally implemented as detailed in \cite{2018arXiv181110454B} to produce the challenge Leaderboards and the results of this paper. The released version$^{2}$ is an independent implementation of the same procedure that improves computational efficiency and portability, while providing consistent results. The scoring procedure evaluates individual frequency submissions and returns the $B_{\nu}$ terms in equation (\ref{gtot}) that contribute to the overall score $G_{\rm tot}$. The FoV$_{\nu}$ terms needed to compute $G_{\rm tot}$ are 30.25, 4.84 and 0.112 square degrees for 560, 1400 and 9200\,MHz respectively.

\section{Results}\label{sec:results}
\begin{table*}
   \caption{Number of detections, matched sources and associated uncertainty (matched sources in the null test) for all the Submitted catalogues.  \label{tab:detections}}
    \centering
    \begin{tabular}{|c|c|c|r|r|r|}
\hline
    Team&Frequency&Depth&$N_{\rm det}$&$N_{\rm m}$&$N_{\rm n}$\\
    \hline
ARCIt-CACAO&560\,MHz&1000h&520166&384778&22298\\%v2020
&1400\,MHz&1000h&150370&143713&3157\\
&9200\,MHz&1000h&765&600&2\\
\hline
&560\,MHz&100h&195935&186982&7450\\
&1400\,MHz&100h&58328&57012&783\\
&9200\,MHz&100h&263&115&2\\
\hline
&560\,MHz&8h&50079&48317&838\\
&1400\,MHz&8h&12468&12222&68\\
&9200\,MHz&8h&235&25&3\\
\hline

\hline
EngageSKA Portugal&560\,MHz&1000h&422038&417909&45601\\%v2
&1400\,MHz&1000h&144147& 142659&43802\\
&9200\,MHz&1000h&633&274&32\\
\hline
\hline
hs&560\,MHz&1000h&39602&1030&569\\
&1400\,MHz&1000h&52932&10654&4130\\
\hline
&560\,MHz&100h&37438&862&207\\
&1400\,MHz&100h&36858&3618&1374\\

\hline
&560\,MHz&8h&19991&241&44\\
&1400\,MHz&8h&12708&677&189\\
\hline
\hline
ICRAR&560\,MHz&1000h&279914&259531&14677\\
&1400\,MHz&1000h&41877&32694&718\\
&9200\,MHz&1000h&733&301&3\\
\hline
\hline
IPM&560\,MHz&1000h&610505&18037&3748\\
&1400\,MHz&1000h&214460&15158&437\\
&9200\,MHz&1000h&16368&696&9\\

\hline
&560\,MHz&8h&83769&5561&132\\
\hline
\hline
IPM2&560\,MHz&1000h&517794&59410&53432\\%v5
&1400\,MHz&1000h&672544&46717&56037\\
&9200\,MHz&1000h&599731&14591&16984\\
\hline
\hline
JLRAT&560\,MHz&1000h&1381466&906914&40779\\ %v5
&1400\,MHz&1000h&391562&249051&18024\\
&9200\,MHz&1000h&6334&2944&18\\
\hline
\hline

RADGK&560\,MHz&1000h&33&3&1\\
&1400\,MHz&1000h&25&4&1\\
\hline
\hline

Shanghai&560\,MHz&1000h&292670&291547&12160\\ %v5
&1400\,MHz&1000h&102129&101687&1727\\
&9200\,MHz&1000h&316&296&1\\
\hline
&560\,MHz&100h&113659&113168&2576\\
&1400\,MHz&100h&32143&31898&308\\
&9200\,MHz&100h&68&60&1\\
\hline
&560\,MHz&8h&21784&21582&192\\
&1400\,MHz&8h&5582&5489&26\\
&9200\,MHz&8h&17&10&1\\
\hline

    \end{tabular}

\end{table*}

Table \ref{tab:detections} lists all the results submitted for SDC1. For each Submitted catalogue, we show the number of detected sources $N_{\rm d}$, the number matches, $N_{\rm m}$, and the number of matches obtained for the Null catalogue, $N_{\rm n}$, providing the estimated number of chance matches in $N_{\rm m}$.

Only two teams (ARCIt-CACAO and Shanghai) submitted entries for all three depths, while the rest of them focused on the 1000\,h exposure. In the following, the 1000\,h results will be considered, unless otherwise specified.

\subsection{Source finding}
From Table \ref{tab:detections} we can observe that the number of detections, $N_{\rm d}$, spans orders of magnitudes between different teams, from some tens of sources detected via visual inspection to some hundred thousands detected with automated pipelines; a similarly large range applies to the number of matches, $N_{\rm m}$. 

Figure \ref{fig:CandR_teams} shows completeness and reliability, defined as: 
\begin{eqnarray}
C(\log F')&=&(N_{\rm m}(\log F' )-N_{\rm n}(\log F'))/N_{\rm t}(\log F') \label{c}\\
R(\log F)&=&(N_{\rm m}(\log F)-N_{\rm n}(\log F))/N_{\rm d}(\log F), \label{r}
\end{eqnarray}
where $F$ is the integrated apparent flux density (before primary beam correction) and the prime denotes the True catalogue value. $N_{\rm t}$, $N_{\rm d}$, $N_{\rm m}$ and $N_{\rm n}$ are the histograms of the True and Submitted catalogues, of their cross-match and of the cross-match obtained for the Null catalogue. We note that $C$ is measured as a function of the True catalogue entries ($\log F’$) and $R$ as a function of the Submitted catalogue entries ($\log F$). This is because the True catalogue and the Submitted catalogue contain only $\log F’$ and $\log F$, respectively. Since the cross-matched catalogues in the numerators contain both $\log F$ and $\log F’$, we computed both histograms in order to achieve consistency with the denominators. This always guarantees the correct normalization of $C$ and $R$ even in the presence of errors in the estimation of the flux $F$.  Whenever such errors are significant, this may however cause some discrepancy between $C$ and $R$. Table \ref{tab:candr} provides a summary of the source-finding performance, in terms of the $C_{\rm tot}$ and $R_{\rm tot}$ metrics, which is consistent with the view provided by Figure \ref{fig:CandR_teams}.

The irregular shapes of the completeness and reliability curves for the 9.2\,GHz channel are due to the much smaller number of sources, which makes these statistics much noisier. At the lowest frequencies, the curves are more regular, although some present interesting features. 
Some of the curves show a drop towards the highest fluxes, which is counter-intuitive. We investigate this aspect below.

\begin{figure}
\includegraphics[width=7.5cm]{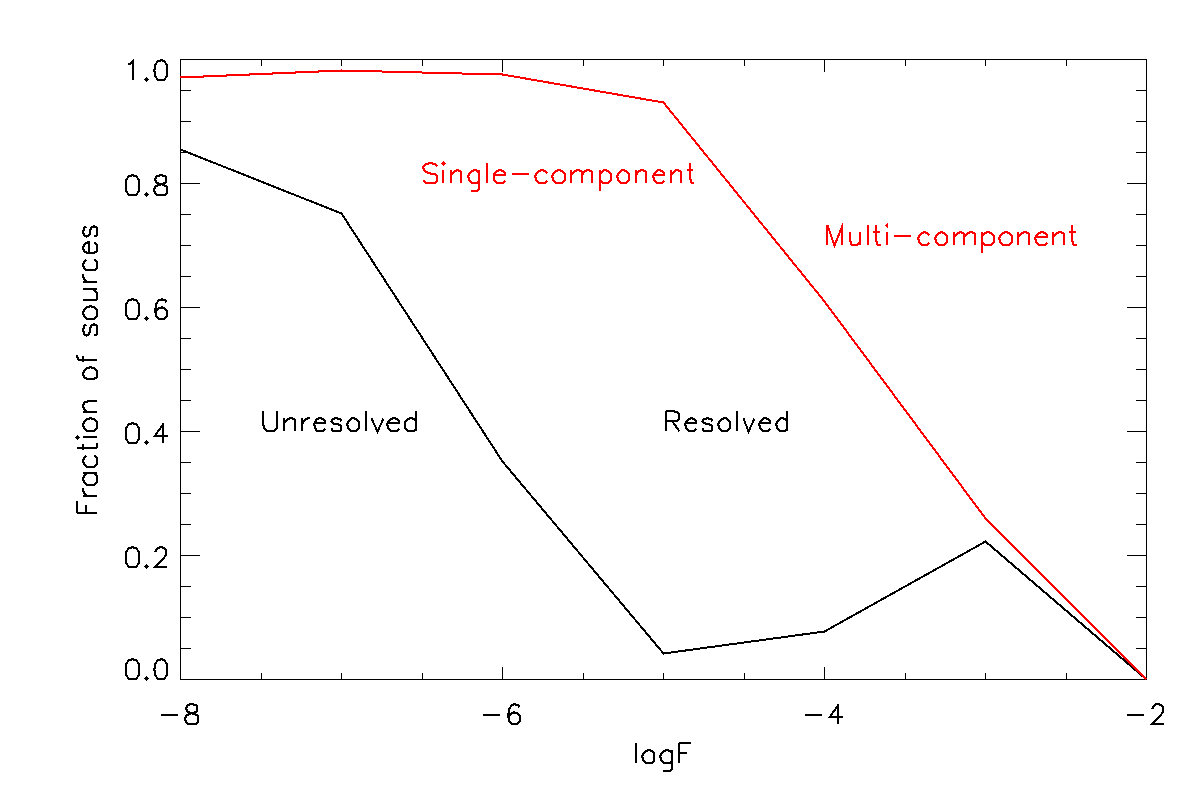}

\caption{Fraction of resolved/rnresolved and single/multi component sources in the SDC1 map at 1.4\,GHz as a function of the apparent flux.}

\label{fig:populations}
\end{figure}

At difference with most previous analyses  \citep[e.g.,][]{hancock2012,hopkins2015,tessa2016},
the SDC1 dataset includes resolved sources and multi-component sources with complex morphology as well as unresolved (point-like) sources. 
In Fig. \ref{fig:populations} we show the fraction of SDC1 sources at 1.4\,GHz in the resolved/unresolved and multi/single component categories as a function of the apparent flux. 
Different categories dominate at different fluxes: 
\begin{itemize}
\item at $F \geq 10^{-3}$\,Jy, the majority of sources are AGN; they typically have a complex morphology and multiple components;
\item for $ 10^{-6} \leq F \leq 10^{-3}$\,Jy the main population is SFGs; they are modelled as single components but they appear still resolved at the 0.6\,arcsec resolution of the 1.4\,GHz map;
\item for $ F \leq 10^{-6}$\,Jy the sources become mostly unresolved. 
\end{itemize}

\begin{figure}
\includegraphics[width=7.5cm]{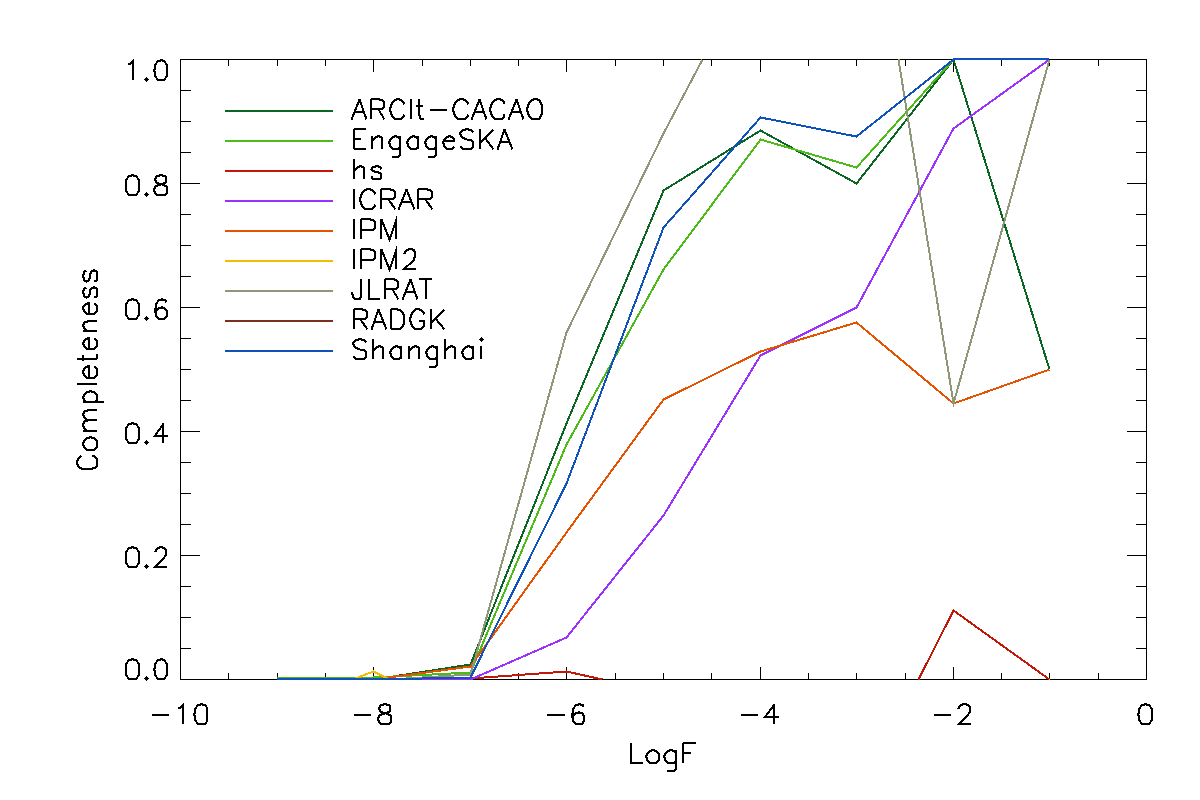}
\includegraphics[width=7.5cm]{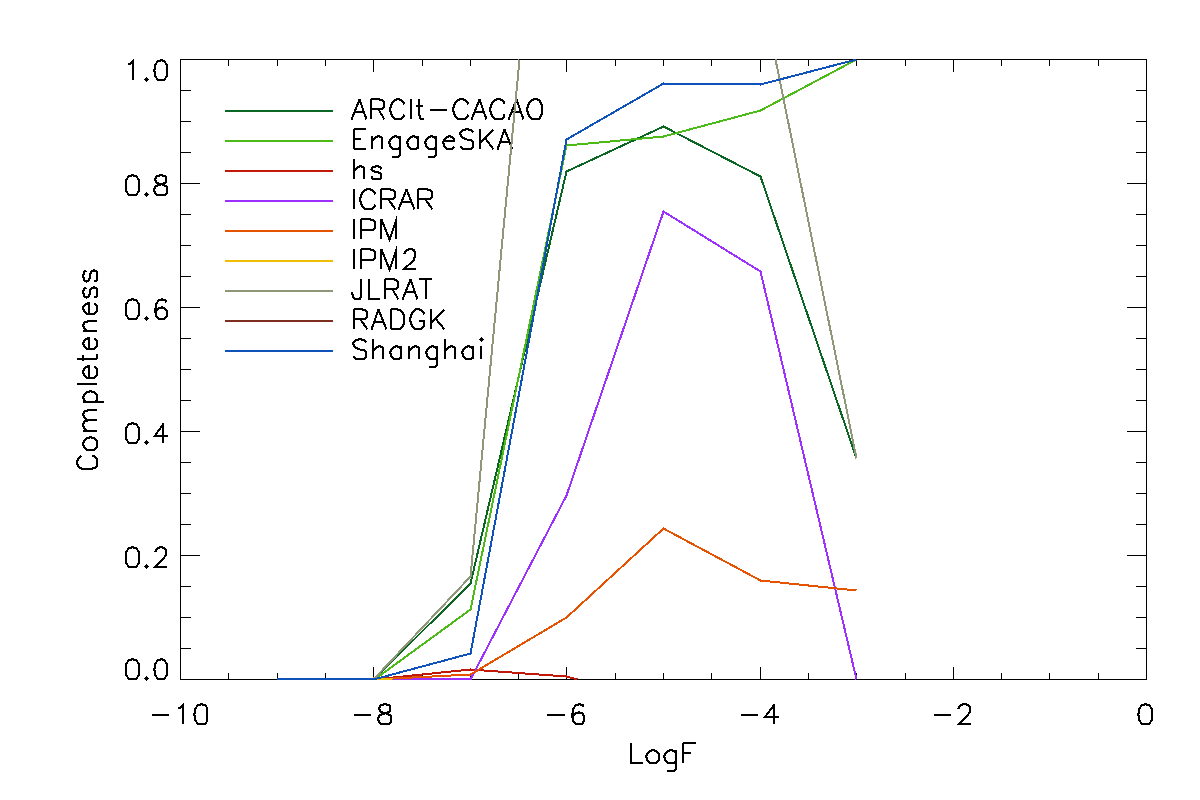}
\caption{Completeness at 1.4\,GHz separately for multi-component (top) and single-component (bottom) sources.}
\label{fig:C4populations}
\end{figure}

In Fig. \ref{fig:C4populations} we show the completeness at 1400\,MHz for the multiple components and the single components separately. Several teams performed quite differently on the two categories, both in terms of shape and of normalization of the curves. A different degree of success of the source-finding methods on the different source categories, dominating at different fluxes, can explain the non-trivial behaviour of completeness and reliability as a function of S/N.

One aspect that played a role in some of these results is the tessellation into smaller sub-images that several teams performed in order to reduce the computational complexity of the analysis (see the tile sizes in Table \ref{tab:teamtessellation}). 
A given tile size would affect source sizes several times smaller, the exact cutoff depending on details of the tessellation strategy (e.g. whether sub-images are overlapping or not) and of the analysis. 
Those sources would either not be identified, or not characrerised well enough; sources consisting of multiple components spanning two tiles would be identified as two independent sources. 
Table \ref{tab:tessellation} shows the total number of sources above a noise level of 5\,$\sigma$ and having Largest Angular Size (LAS) above given sizes. These numbers can be compared to those in Tables \ref{tab:intcounts} and \ref{tab:detections} to have an idea of the fraction of sources potentially missed due to tessellation.  

We can see that tile sizes of some hundreds of pixels would reduce the number of detections, with some impact on performance possibly happening up to a tile size of around 1000 pixels. ARCIt-CACAO, adopting a 1024 pixels tile, ascribes the features in Fig. \ref{fig:C4populations} to tessellation effects (see section \ref{sec:cacao} for more details).  

\begin{table}
   \caption{Size of the tiles used by the teams to divide the original SDC1 images for source detection, ordered by increasing size.}
    \centering
    \begin{tabular}{l|r}
&Side of square\\
Teams& tile (pixels)\\
\hline
hs&50\\
ICRAR&205\\
JLRAT&320\\
IPM2&1000\\
ARCIt-CACAO&1024\\
EngageSKA&16384\\
IPM& full size\\
RADGK&full size\\
Shanghai&full size\\
\end{tabular}
\label{tab:teamtessellation}
\end{table}

\begin{table}
   \caption{Number of $>5\,\sigma$ sources with LAS above given sizes in pixels.}
    \centering
    \begin{tabular}{r|r|r|r}
LAS (pixels)&560\,MHz&1400\,MHz&9200\,MHz\\
\hline
25&6561&6555&114\\
50&1147&1369&45\\
100&98&107&7\\
200&20&30&6\\
500&3&6&2\\
1000&1&2&1\\
15000&1&1&1\\
\end{tabular}
\label{tab:tessellation}
\end{table}

Overall, our analysis supports the picture that the higher level of realism in the morphology of the SDC1 sources is responsible for some reduction in performance of the source-finding methods with respect to the ideal case. 
This is something to bear in mind as future high-resolution observations will expose the full complexity of the real sky. When needed for computational reasons, any tessellation of the field of view should be designed to cope with the presence of extended and multi-component sources, by choosing carefully the tile sizes or by performing the analysis at multiple scales and spatial resolutions.

\begin{figure}

\includegraphics[width=7.5cm]{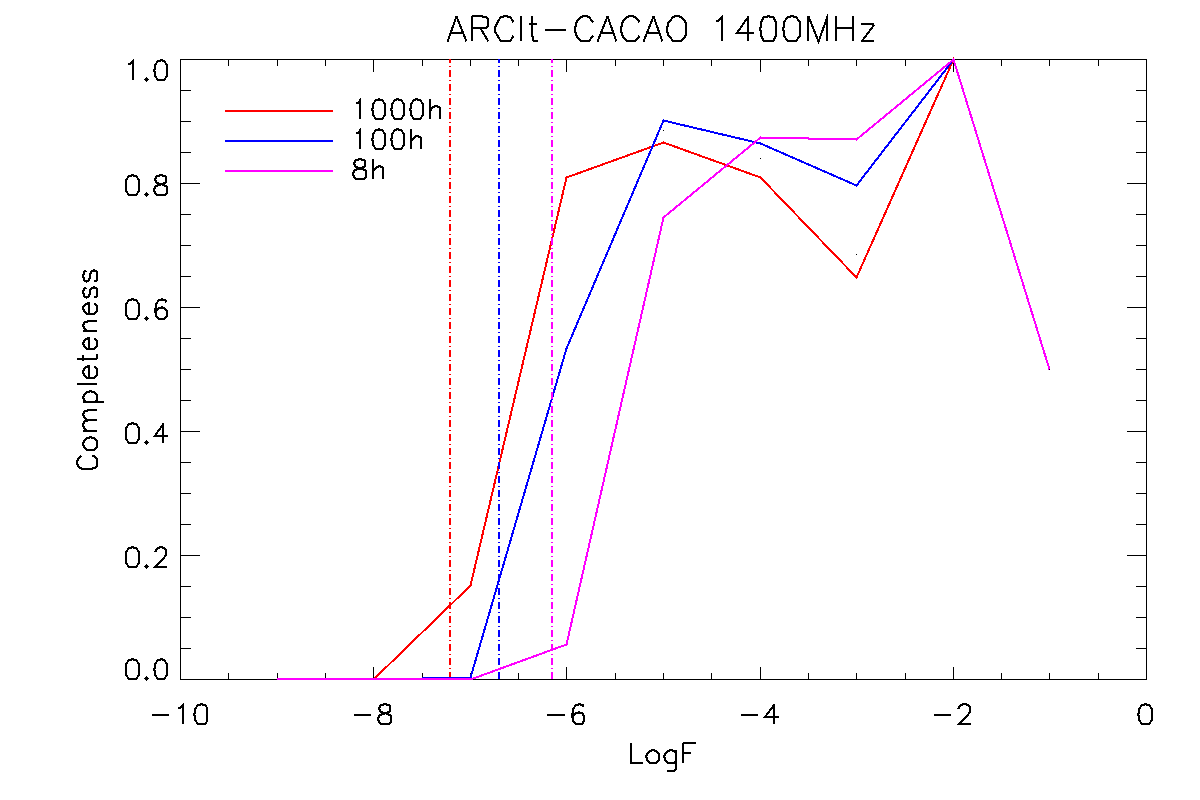}
\includegraphics[width=7.5cm]{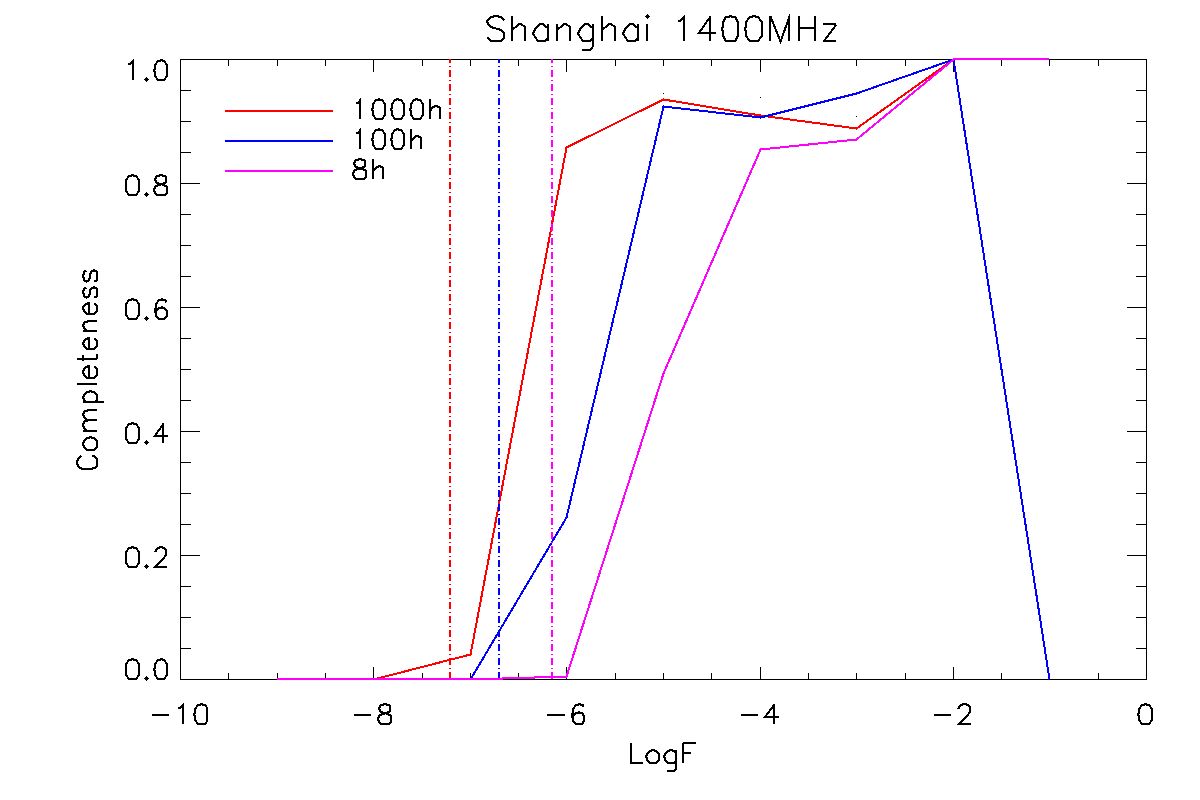}

\caption{Completeness as a function of depth for the ARCIt-CACAO (top) and Shanghai (bottom) teams at 1.4\,GHz. The vertical dotted lines are the 1\,$\sigma$ noise RMS.}
\label{fig:C_depth}
\end{figure}

\begin{table}

    \centering
       \caption{$C_{\rm tot}$ and $R_{\rm tot}$ metrics for all teams, in order of decreasing $C_{\rm tot}$ \label{tab:candr}}
    \begin{tabular}{l|c|c}
         Teams&$C_{\rm tot}$&$R_{\rm tot}$        \\
\hline
IPM2 &     141801   &  0.07  \\
JLRAT &     107705  &   0.58  \\
ARCIt-CACAO &  47766.1 & 0.83    \\%v2020
EngageSKA   &   45734.9   &  0.80 \\
Shanghai    &  33288.6 &   0.98\\ 
ICRAR    &  18020.1 &     0.71 \\
IPM &     9937.96   &  0.05  \\
hs    &  2235.29    &0.07  \\
RADGK    & 0.9256   & 0.08 \\
    \end{tabular}
 
    \end{table}

Figure \ref{fig:C_depth} shows how Completeness at 1400\,MHz varies with varying depth, from 1000\,h to 8\,h, for the three teams providing those catalogues. All those teams managed to improve their catalogues for a deeper image, which is indicative of a noise-limited rather than a systematics-limited performance. The 50\% completeness for both the 100\,h and the 1000\,h exposure is at the 4.5--14\,$\sigma$ depending on the teams.

\subsection{Source characterization}
Figure \ref{fig:Ddistr} shows the distribution of the errors on all source attributes for all teams. Errors refer to the 560\,MHz submission, which provides the largest source statistics, however the results at other frequencies are similar. By definition all these errors are positive and therefore the distributions are asymmetric. In all cases, a good performance produces a narrow distribution (small random error) peaked on 0 (no bias).  

The left panels show the positional, flux and size errors $D_{\rm pos}$, $D_{\rm size}$ and $D_{\rm flux}$ as defined in eqns. (\ref{dpos})--(\ref{dflux}). These error components are particularly important because of their role in the cross-match procedure. If their sum in quadrature $D$ [eq. (\ref{d})] exceeds the threshold value, the source is discarded from further analysis and classified as a false positive. Some teams have a large random or systematic error on one or more of those attributes, and this is at the root of low scores in Table \ref{tab:candr}. 

Position is generally well recovered, with only a few exceptions. Error on size is generally wider, which reflects the different size definition adopted by different approaches and difficulty in deconvolving the beam size. The flux error varies between different teams from very small (ARCIt-CACAO, ICRAR, Shanghai) to quite large (hs, IPM).

The right panels show the position angle and core fraction errors $D_{\rm PA}$ and $D_{\rm cf}$ and $D_{\rm class}$ as defined in eqns. (\ref{dcf})--(\ref{dclass}) and the accuracy of classification. 

Most $D_{\rm PA}$ distributions are relatively flat, indicating this attribute is not very successfully recovered. 
Some of the distributions for the core fraction error are bimodal because the true core fraction distribution is bimodal ($\rm{cf}=0$ for star-forming galaxies and $\rm{cf} \neq 0$ for AGN). 
The accuracy of source classification is discussed in the next sub-section. 

In Table \ref{tab:ganda} we show the $G_{\rm tot}$ and $A_{\rm tot}$ scores for all teams, in order of $G_{\rm tot}$. The accuracy score $A_{\rm tot}$ provides a good summary of the performance shown in Figure \ref{fig:Ddistr}. The global score $G_{\rm tot}$ summarizes all metrics previously introduced, and $G^*_{\rm tot}$ records its value as it was achieved by the challenge deadline (the SDC1 leaderboard is based on $G^*_{\rm tot}$).

\subsection{Source classification}

Overall source classification was not probed well enough by this challenge. The simulated observing strategy (a single telescope pointing) did not provide enough multi-frequency information for the sources, which could have helped classification. 
Similarly, the challenge did not include data at other wavelengths, most notably optical and/or IR, which is often used to separate star-forming galaxies from AGNs by means of the methods involving optical emission lines information such as BPT diagram and radio-IR correlation.
These issues, inherent in the challenge design, are now understood and will inform the development of future data challenges featuring a source classification aspect. Since classification contributes only marginally to the final score, the impact of this design issue on the overall challenge is limited.

\begin{figure*}
\includegraphics[width=8cm]{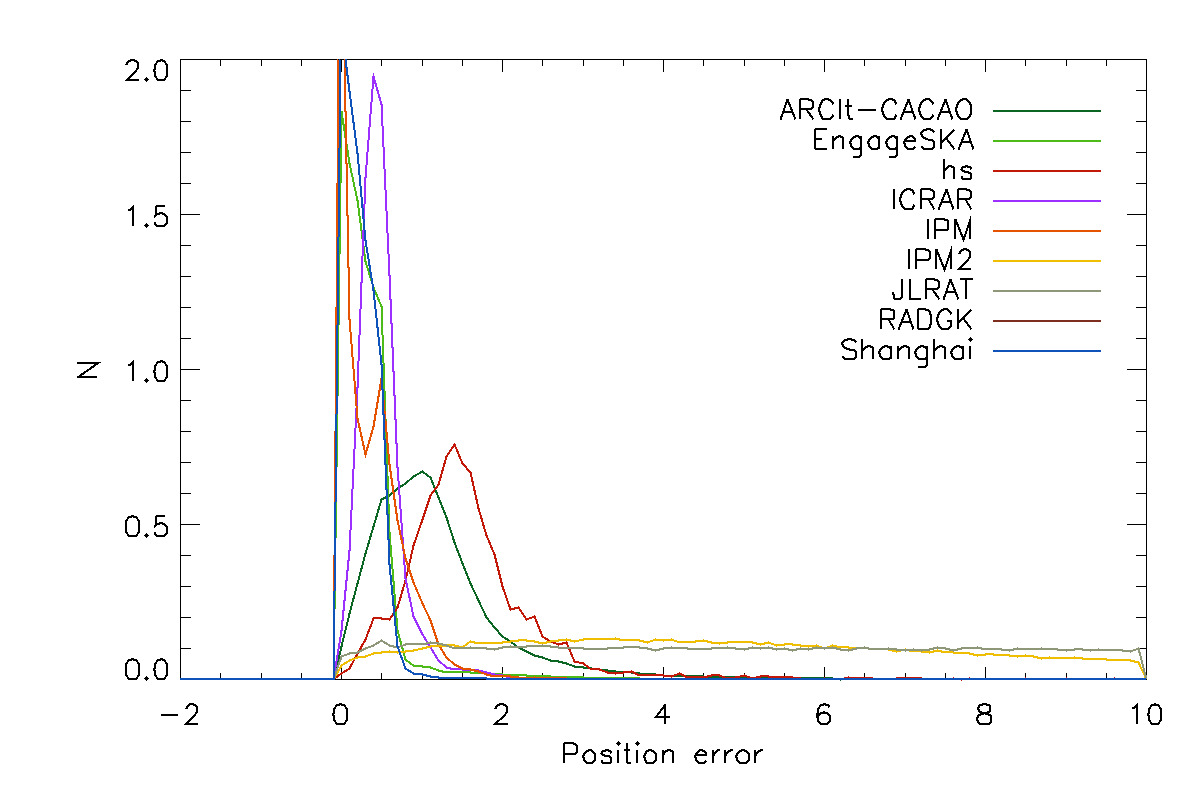}
\includegraphics[width=8cm]{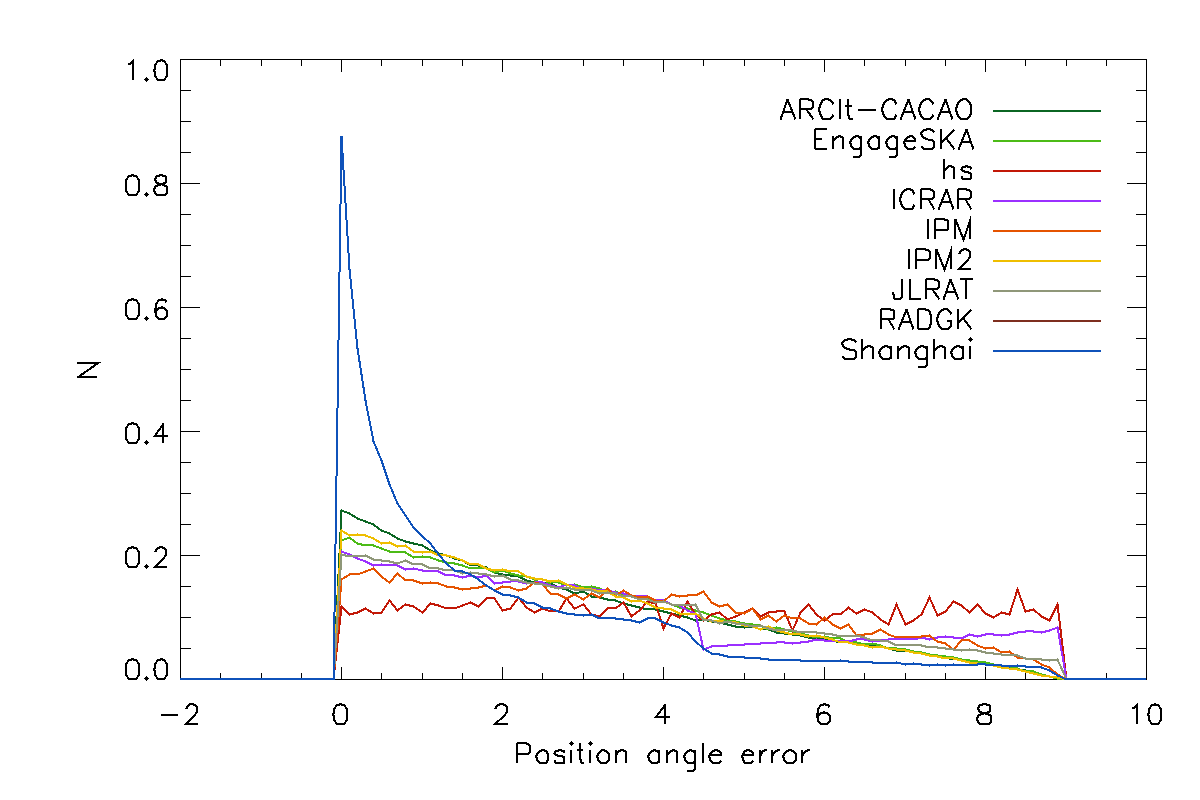}

\includegraphics[width=8cm]{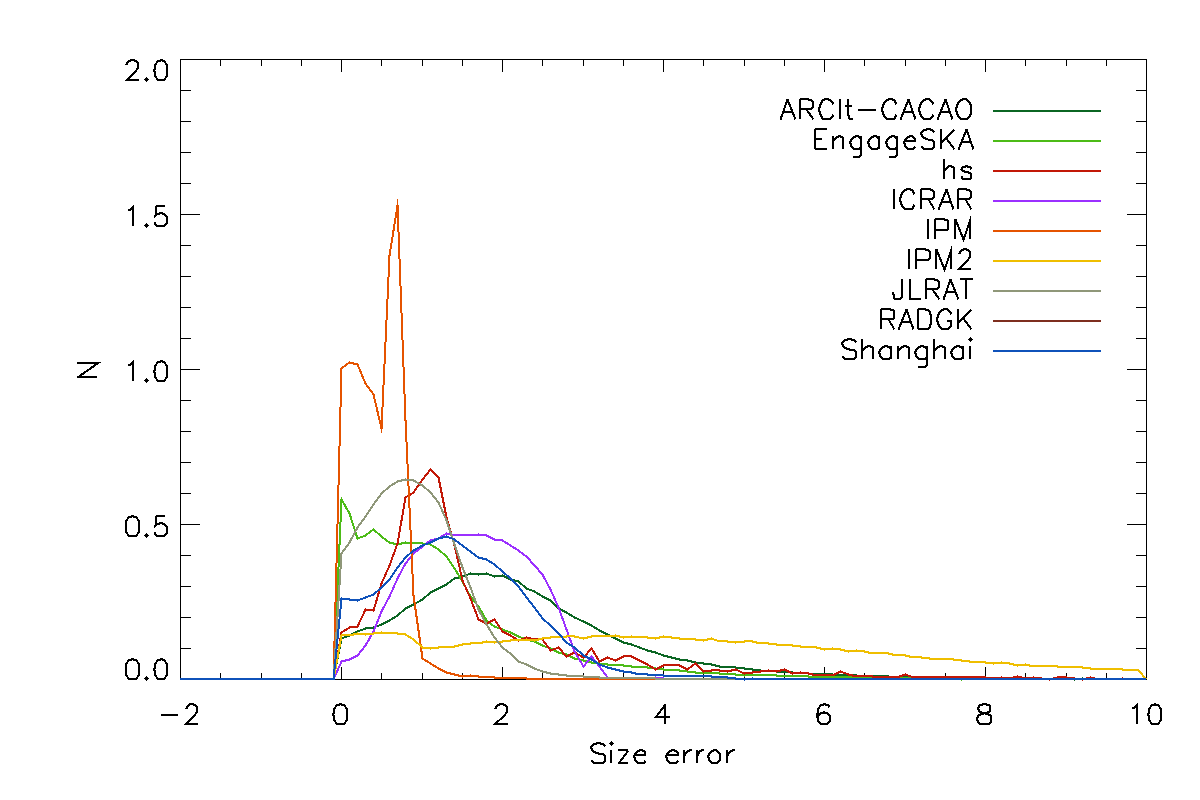}
\includegraphics[width=8cm]{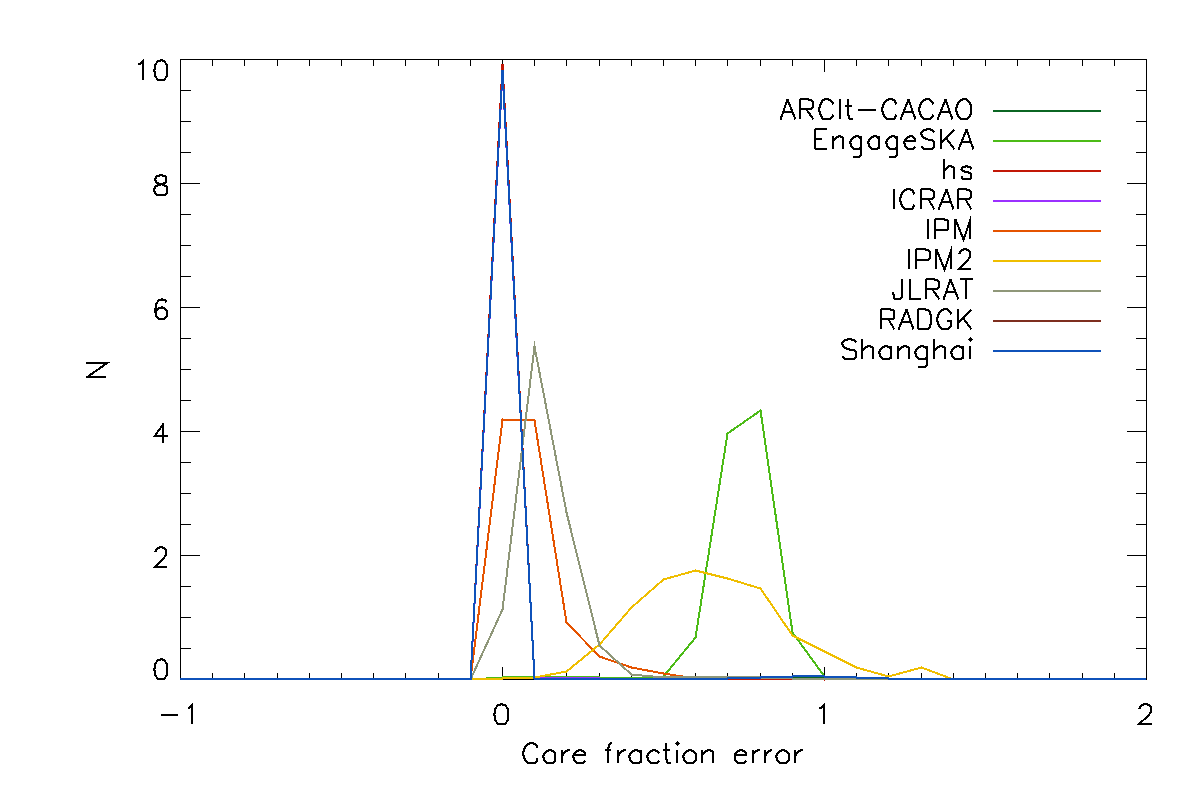}

\includegraphics[width=8cm]{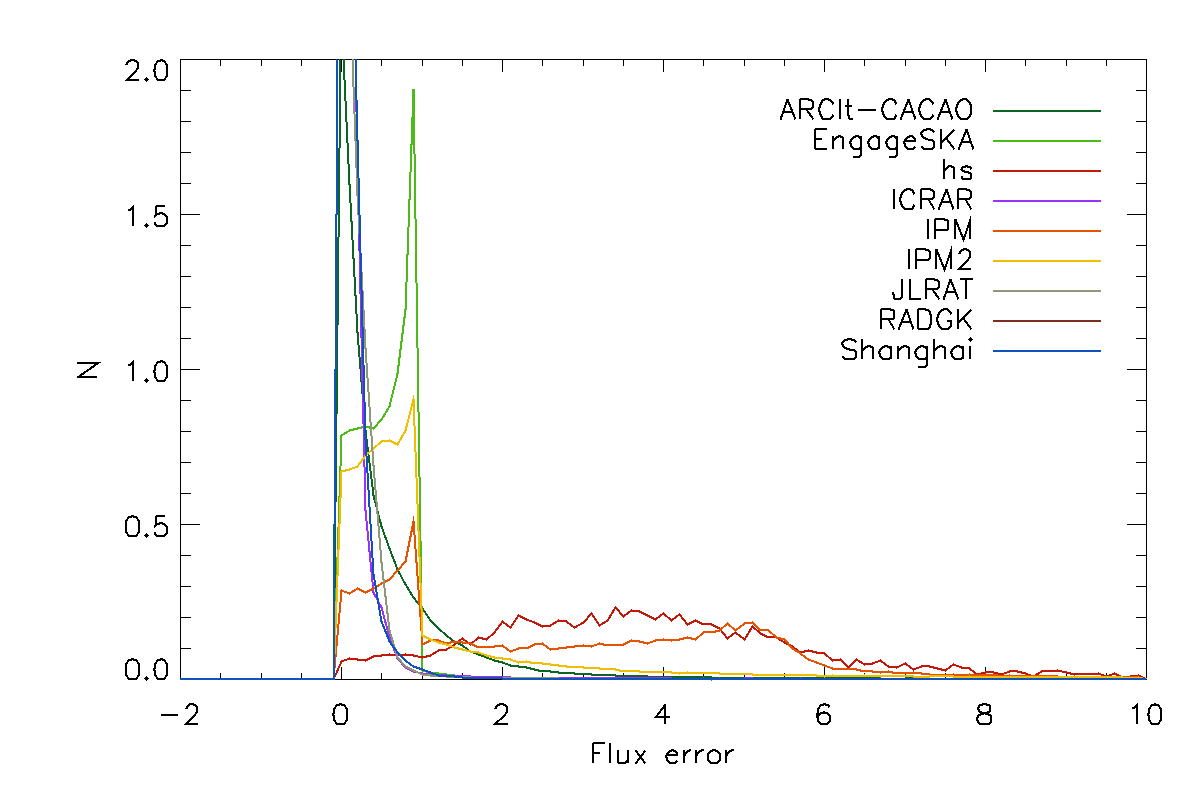}
\includegraphics[width=8cm]{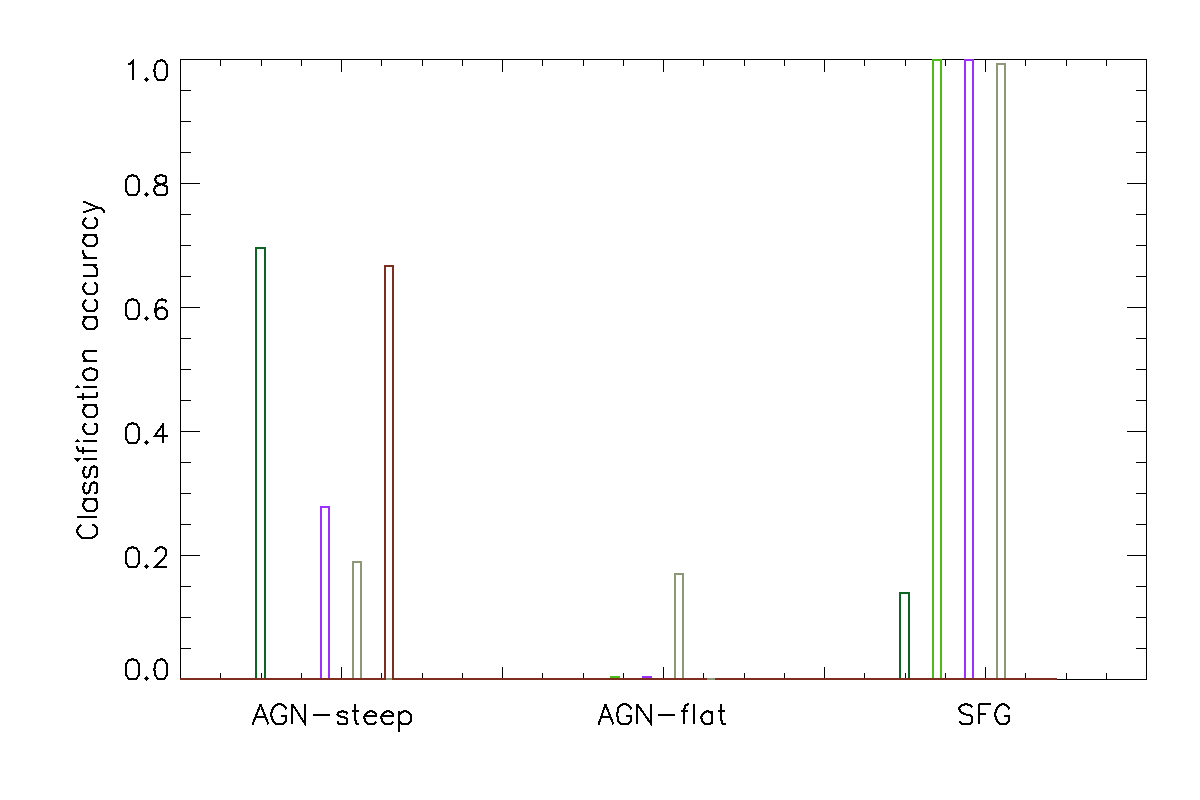}
\caption{Distribution of errors on the source attributes for all teams on the 560\,MHz 1000\,h submissions.}

\label{fig:Ddistr}
\end{figure*}

ARCIt-CACAO, EngageSKA-Portugal, ICRAR, JLRAT and RADGK attempted source classification.  In Table \ref{tab:class} we report how many sources have been classified in total by each team at 560\,MHz, and how the true classes (on different columns) have been distributed into the estimated classes (in different rows). The perfect classification would correspond to all objects being in the diagonal of the first three rows. To visualise some of this information, the bottom right panel of Fig. \ref{fig:Ddistr} shows the fraction of correctly identified sources (on the diagonal) of each true class (sum of all elements of a column) for AGN-steep, AGN-flat and SFGs respectively.

EngageSKA, ICRAR and JLRAT have almost all SFGs correctly identified. Their strategy has been to classify sources initially as SFGs, since they are the outstanding majority of the total, and to update their class to AGN if that was constrained by the data. This strategy maximises the total score $G_{\rm tot}$ for their submission. ARCIt-CACAO did not attribute a class to all sources in their submitted catalogue, therefore they have a lower percentage of correctly identified SFGs relative to the total (around 20\,\%).

Around 70\,\% of steep-spectrum AGN were reliably classified by ARCIt-CACAO and, albeit on a very small sample, RADGK. All teams struggled with the classification of the flat-spectrum AGN. The best results here is a 20\,\% correct classification by JLRAT. Flat-spectrum AGN are compact objects, so they cannot easily be morphologically distinguished from SFGs. 
Due to the simulated observing strategy of a single telescope pointing, the full spectral information, that could have helped classification, was available only for the small number of sources detectable at 9.2\,GHz.

\begin{table}

    \centering
       \caption{$G_{\rm tot}$ and $A_{\rm tot}$ metrics for all teams, in order of decreasing $G_{\rm tot}$. $G^*_{\rm tot}$ is the total score achieved by the deadline of 30$^{\rm th}$ April 2019, which determined the SDC1 leaderboard$^7$. \label{tab:ganda}}
    \begin{tabular}{l|c|c|c}
         Teams&$G_{\rm tot}$&$G^*_{\rm tot}$&$A_{\rm tot}$        \\
\hline
Shanghai    & 19112.8  & -33226.1 & 19419.7  \\ 
ARCIt-CACAO & 17361.3    & 2733.58 & 24684.6     \\
EngageSKA   &  16914.9   & 16914.9& 20551.5  \\
ICRAR    &  5265.56  & 5265.56 &11691.1  \\
RADGK    &  -4.58427 & -4.58427&0.746315 \\
hs    &  -9325.29   & -9325.29 &684.933 \\
JLRAT &  -10625.9   &  -53069.4 &64752.6    \\
IPM &  -196237     & -196237  &4356.57\\
IPM2 &  -533625     & - & 28973.2\\
    \end{tabular}
 
    \end{table}
\begin{table}

    \centering
       \caption{Source classification statistics. The columns are the true classes and the rows the estimated classes for all teams that performed classification \label{tab:class}}
    \begin{tabular}{l|l|r|r|r}
&&AGN-steep&AGN-flat&SFG\\
\hline

ARCIt-CACAO&AGN-steep&5858&1972&181325\\
&AGN-flat&0&0&0\\
&SFG&1745&1629&51640\\
&none&826&1656&138158\\
\hline
EngageSKA&AGN-steep&6&23&64\\
&AGN-flat&2&18&369\\
&SFG&8340&5780&403547\\
&none&0&0&0\\
\hline
ICRAR&AGN-steep&1266&6&309\\
&AGN-flat&4&13&1\\
&SFG&3293&4233&261066\\
&none&0&0&0\\
\hline
JLRAT&AGN-steep&3066&395&3552\\
&AGN-flat&394&3236&6543\\
&SFG&12737&15471&1268354\\
&none&0&0&0\\
\hline
RADGK&AGN-steep&2&0&1\\
&AGN-flat&1&0&1\\
&SFG&0&0&0\\
&none&0&0&0\\
\end{tabular}
\end{table}

\section{Conclusions and discussion} \label{sec:conclusions}

With the SKA science data challenge 1 (SDC1) we started investigating the properties of SKA continuum imaging products and the issues associated with their analysis.
The challenge was meant as a training opportunity, to develop the skills of the astronomical community in the analysis of SKA-like data. Different teams approached it from a different level of expertise in the specific kind of analysis, and a different level of readiness in terms of existing pipelines. In all cases the challenge constituted a step forward in the understanding of the problem and in specific expertise, which in itself is a huge success. 
The conclusions from this analysis are as follows:

\begin{itemize}
\item Eight independent teams, using various approaches (described in Sec. \ref{sec:methods}), succeeded in processing the images for source detection, classification and characterization. 
\item The high spatial density of sources challenged the standard procedure of confirming source detections through a positional cross-match. Additional source properties (flux, size) were used to distinguish between multiple matches consistent with each source position. The same issue will likely be encountered with real data as the depth of surveys increase.
\item Several teams achieved good performance in the source finding and/or source characterization aspect of the challenge and showed complementary strengths and weaknesses, which highlights the importance of applying multiple pipelines to the data.
\item The size of the data (4 GB per map) was just a first step towards the full SKA complexity, with a prospect to go further in future exercises. It has been overcome by most teams by splitting the data into sub-regions and analysing them separately. In some cases the size of those regions was too small to successfully detect and characterise the more extended sources, with some impact on the performance. 
\item Teams who delivered results for different depths (8\,h, 100\,h, 1000\,h) succeeded in getting significantly higher detection rates for longer exposures. At 1.4\,GHz, 50\% completeness varies across those teams between 4.5 and 14\,$\sigma$. Most notably, it is stable going from the 100\,h to the 1000\,h exposure, which indicates that the challenge due to source crowding has been overcome very well.  At 560\,MHz, consistently with the lower spatial resolution, crowding might be playing a somewhat bigger role, as the 50\% completeness thresholds double from 5 to 10\,$\sigma$ going from medium to long integration. The very good image quality of the SDC1 simulation might have given an optimistic view on the subject of source crowding, which should be reassessed after introducing calibration, pointing and deconvolution errors. 
 \item Some teams had difficulties in setting an optimal signal-to-noise threshold for source detection, which resulted in either a low completeness or a low reliability. In this respect, the specific expertise within teams in this kind of analysis played a major role. This stresses the importance of building domain knowledge and expertise alongside developing good, publicly available software for radio data analysis. 
  \item The increased realism of the SDC1 simulation (particularly the presence of resolved sources and complex morphology with multiple components) meant a decrease in performance of source-finding methods with respect to the ideal case, where completeness and reliability approach 100\,\% at the highest S/N. Various approaches were particularly successful on one category of sources, but none was consistently good across all source types. This is something to bear in mind as future high-resolution observations will expose the full complexity of the real sky. 
\item Accurate integrated flux and size estimation presented some problems. Both aspects are again linked to the complexity and the diversity of the source morphology, and to attempts by the adopted methods to fit them all with a single model. For example, Gaussian fitting is adequate on point sources but less so on resolved ones. Within SDC1, we allowed three alternative size definitions to fully represent the range of morphologies injected. Every conversion from those definitions to the ones used by the participating teams is just an approximation and can lead to a biased result. 
\item The source classification (into SFG, steep-spectrum and flat-spectrum AGN) was not probed well enough by this challenge. This specific aspect deserves further investigation, possibly by means of dedicated exercises.
\end{itemize}

\section*{Data availability}

\begin{itemize}
\item The SDC1 dataset is available on \url{https://astronomers.skatelescope.org/ ska- science- data- challenge- 1/}
\item The SDC1 scoring package is available on \url{https://pypi.org/project/ska- sdc/}
\item Results from the ICRAR team are available on \url{https://github.com/ICRAR/skasdc1}
\item Most of the codes  used by the participating teams are available publicly (see Sec \ref{sec:methods} for more details).
\end{itemize}

\section*{Acknowledgements}
The authors thank the anonymous referee for useful comments and suggestions, that improved the paper.
\emph{ARCIt-CACAO}: This research made use of Astropy,\footnote{\url{http://www.astropy.org}} a community-developed core Python package for Astronomy \citep{astropy:2013, astropy:2018}.

\emph{EngageSKA-Portugal}: The Enabling Green E-science for the Square Kilometre Array Research Infrastructure (ENGAGE-SKA) team acknowledges financial support from grant POCI-01-0145-FEDER022217, funded by Programa Operacional Competitividade e Internacionaliza\c{c}\~ao (COMPETE 2020) and the Funda\c{c}\~ao para a Ci\^encia e a Tecnologia (FCT), Portugal. This work was also funded by FCT and Minist\'erio da Ci\^encia, Tecnologia e Ensino Superior (MCTES) through national funds and when applicable co-funded EU funds under the project UIDB/50008/2020-UIDP/50008/2020 and UID/EEA/50008/2019. B.C. acknowledges support from the Advanced EU Network of E-infrastructures for Astronomy with SKA (AENEAS), funded by the European Commission Framework Programme Horizon 2020 RIA under grant agreement n. 731016.  S.A. acknowledges support by FCT, through CIDMA, within project UIDB/04106/2020. V.A.R.M.R. acknowledges support from the FCT Investigator through exploratory project of reference IF/00498/2015/CP1302/CT0001 and Origin, composition, evolution and exploration of Phobos (PHOBOS), project reference POCI-01-0145-FEDER-029932, funded by COMPETE 2020 and FCT, Portugal.

\emph{hs}: MB acknowledges support from the Deutsche Forschungsgemeinschaft under Germany's Excellence Strategy - EXC 2121 "Quantum Universe" - 390833306.

\emph{ICRAR}: The ICRAR team acknowledges the use of CSIRO/Bracewell, ICRAR/Pleiades.

\emph{IPM}: The IPM team acknowledges the School of Astronomy's SKA group led by F. Tabatabaei for initiating this challenge and guidance.
It also thanks H. Khosroshahi for his help with the SExtractror usage. The IPM team is thankful to J. Miraftabzadeh and H. Hassani for their supports. In addition, IPM2 team thanks F. Arefi for his useful hints.

\emph{JLRAT}:  This work is supported by the National Key R\&D Program of China under grant number 2018YFA0404703 and the Open Project Program of the Key Laboratory of FAST, NAOC, Chinese Academy of Sciences.

\emph{Shanghai}: The Shanghai Astronomical Observatory team acknowledges the funding support from the National Key R\&D Programme of China (2018YFA0404603). The data processing is carried out on the China SKA Regional Centre prototype \citep{2019NatAs...3.1030A}. 

\emph{SKAO}: The SKA organization acknowledges the use of the facilities of the Italian Center for Astronomical Archive (IA2) operated by INAF to store the SDC1 dataset. We thank the Ox-ICRAR team (M. Jarvis, C. Hale, A. Robotham) for their contribution to the challenge and useful comments on this paper.

\section*{Affiliations}
$^1$ SKA Organization, Jodrell Bank, Lower Whitington, Macclesfield, SK11 9DL, UK\\
$^2$ Jodrell Bank Centre for Astrophysics, Department of Physics \& Astronomy, The University of Manchester, Manchester M13 9PL, UK\\
$^3$ Shanghai Astronomical Observatory, Key Laboratory of Radio Astronomy, Chinese Academy of Sciences, 80 Nandan Road, Shanghai 200030, China\\
$^4$ Hamburger Sternwarte, University of Hamburg, Gojenbergsweg 112, 21029 Hamburg, Germany\\
$^5$ INAF, Istituto di Radioastronomia, Italian ARC, Via P. Gobetti 101, Bologna, Italy\\
$^6$ Instituto de Telecomunicações, Campus Universitário de Santiago, 3810-193 Aveiro, Portugal\\
$^7$ School of Astronomy, Institute for Research in Fundamental Sciences (IPM), P.O. Box 1956836613, Tehran, Iran\\%IPM
$^{8}$ Department of Physics, Indian Institute of Technology Kanpur, Uttar Pradesh-208016, India\\%RADGK
$^{9}$ ICRAR-M468, UWA, 35 Stirling Hwy, Crawley, WA 6009, Australia\\% (Email: chenwuperth@gmail.com)\\
$^{10}$CAS Key Laboratory of FAST, National Astronomical Observatories, Chinese Academy of Sciences, Beijing, China\\
$^{11}$CIDMA, Departamento de Física, Universidade de Aveiro, Campus Universitário de Santiago, 3810-193 Aveiro, Portugal\\
$^{12}$Research Institute for Astronomy and Astrophysics of Maraghe, Maraghe, Iran\\
$^{13}$Universidade de Aveiro, Campus Universitário de Santiago, 3810-193 Aveiro, Portugal\\
$^{14}$CICGE, Faculdade de Ciências da Universidade do Porto, Observatório Astronómico, Alameda do Monte da Virgem, 4430-146 Vila Nova de Gaia, Portugal\\
$^{15}$Departamento de Física, Universidade de Aveiro, Campus Universitário de Santiago, 3810-193 Aveiro, Portugal\\
$^{16}$INAF–IAPS, Via Fosso del Cavaliere 100, Rome, Italy\\
$^{17}$Department of Physics, Institute for Advanced Studies in Basic Sciences (IASBS), P.O. Box 11365-9161, Zanjan, Iran\\
$^{18}$School of Physics, Institute for Research in Fundamental Sciences (IPM), P. O. Box 19395-5531, Tehran, Iran\\
$^{19}$Department of Systems Design Engineering, University of Waterloo, Waterloo, Ontario, Canada\\
$^{20}$ ARC Centre of Excellence for Astrophysics in Three Dimensions (ASTRO 3D), Australia\\
$^{21}$ CSIRO Astronomy \& Space Science, PO Box 1130, Bentley, WA 6102, Australia\\

\bibliography{sdc1biblio}

\begin{thebibliography}{55}
\providecommand{\natexlab}[1]{#1}
\providecommand{\url}[1]{\texttt{#1}}
\providecommand{\urlprefix}{URL }
\providecommand{\eprint}[1][]{\url{#1}}

\bibitem[{{An} et~al.(2019){An}, {Wu} \& {Hong}}]{2019NatAs...3.1030A}
{An}, T., {Wu}, X.-P., {Hong}, X., 2019, Nature Astronomy, 3, 1030

\bibitem[{{Astropy Collaboration} et~al.(2018){Astropy Collaboration},
  {Price-Whelan}, {Sip{\H{o}}cz} et~al.}]{Astropy2018}
{Astropy Collaboration}, {Price-Whelan}, A.~M., {Sip{\H{o}}cz}, B.~M., et~al.,
  2018, \aj, 156, 3, 123, \eprint arXiv:{1801.02634}

\bibitem[{{Astropy Collaboration} et~al.(2013){Astropy Collaboration},
  {Robitaille}, {Tollerud} et~al.}]{Astropy2013}
{Astropy Collaboration}, {Robitaille}, T.~P., {Tollerud}, E.~J., et~al., 2013,
  \aap, 558, A33, \eprint arXiv:{1307.6212}

\bibitem[{{Banfield} et~al.(2016){Banfield}, {Andernach}, {Kapi{\'n}ska}
  et~al.}]{banfield16}
{Banfield}, J.~K., {Andernach}, H., {Kapi{\'n}ska}, A.~D., et~al., 2016,
  \mnras, 460, 3, 2376, \eprint arXiv:{1606.05016}

\bibitem[{{Banfield} et~al.(2015){Banfield}, {Wong}, {Willett}
  et~al.}]{banfield15}
{Banfield}, J.~K., {Wong}, O.~I., {Willett}, K.~W., et~al., 2015, \mnras, 453,
  3, 2326, \eprint arXiv:{1507.07272}

\bibitem[{{Becker} et~al.(1995){Becker}, {White} \& {Helfand}}]{becker95}
{Becker}, R.~H., {White}, R.~L., {Helfand}, D.~J., 1995, \apj, 450, 559

\bibitem[{{Bertin} \& {Arnouts}(1996)}]{1996A&AS..117..393B}
{Bertin}, E., {Arnouts}, S., 1996, \aaps, 117, 393

\bibitem[{{Bonaldi} et~al.(2019){Bonaldi}, {Bonato}, {Galluzzi}
  et~al.}]{2019MNRAS.482....2B}
{Bonaldi}, A., {Bonato}, M., {Galluzzi}, V., et~al., 2019, \mnras, 482, 1, 2,
  \eprint arXiv:{1805.05222}

\bibitem[{{Bonaldi} \& {Braun}(2018)}]{2018arXiv181110454B}
{Bonaldi}, A., {Braun}, R., 2018, arXiv e-prints, arXiv:1811.10454, \eprint
  arXiv:{1811.10454}

\bibitem[{{Bonnarel} et~al.(2000){Bonnarel}, {Fernique}, {Bienaym{\'e}}
  et~al.}]{2000A&AS..143...33B}
{Bonnarel}, F., {Fernique}, P., {Bienaym{\'e}}, O., et~al., 2000, \aaps, 143,
  33

\bibitem[{Cai et~al.(2016)Cai, Fan, Feris \& Vasconcelos}]{Cai2016AUM}
Cai, Z., Fan, Q., Feris, R.~S., Vasconcelos, N., 2016, ArXiv, abs/1607.07155

\bibitem[{{Condon} et~al.(2012){Condon}, {Cotton}, {Fomalont}
  et~al.}]{2012ApJ...758...23C}
{Condon}, J.~J., {Cotton}, W.~D., {Fomalont}, E.~B., et~al., 2012, ApJ, 758, 1,
  23, \eprint arXiv:{1207.2439}

\bibitem[{{Frean} et~al.(2014){Frean}, {Friedlander}, {Johnston-Hollitt} \&
  {Hollitt}}]{2014AIPC.1636...55F}
{Frean}, M., {Friedlander}, A., {Johnston-Hollitt}, M., {Hollitt}, C., 2014, in
  Bayesian Inference and Maximum Entropy Methods in Science and Engineering,
  vol. 1636 of \emph{American Institute of Physics Conference Series}, 55--61

\bibitem[{{Hale} et~al.(2019){Hale}, {Robotham}, {Davies}, {Jarvis}, {Driver}
  \& {Heywood}}]{Hale2019}
{Hale}, C.~L., {Robotham}, A.~S.~G., {Davies}, L.~J.~M., {Jarvis}, M.~J.,
  {Driver}, S.~P., {Heywood}, I., 2019, \mnras, 487, 3, 3971, \eprint
  arXiv:{1902.01440}

\bibitem[{{Hales} et~al.(2012){Hales}, {Murphy}, {Curran}, {Middelberg},
  {Gaensler} \& {Norris}}]{2012MNRAS.425..979H}
{Hales}, C.~A., {Murphy}, T., {Curran}, J.~R., {Middelberg}, E., {Gaensler},
  B.~M., {Norris}, R.~P., 2012, \mnras, 425, 2, 979, \eprint arXiv:{1205.5313}

\bibitem[{{Hancock} et~al.(2012){Hancock}, {Murphy}, {Gaensler}, {Hopkins} \&
  {Curran}}]{hancock2012}
{Hancock}, P.~J., {Murphy}, T., {Gaensler}, B.~M., {Hopkins}, A., {Curran},
  J.~R., 2012, \mnras, 422, 2, 1812, \eprint arXiv:{1202.4500}

\bibitem[{{Hancock} et~al.(2018){Hancock}, {Trott} \&
  {Hurley-Walker}}]{2018PASA...35...11H}
{Hancock}, P.~J., {Trott}, C.~M., {Hurley-Walker}, N., 2018, \pasa, 35, e011,
  \eprint arXiv:{1801.05548}

\bibitem[{{Harrison} et~al.(2020){Harrison}, {Brown}, {Tunbridge}
  et~al.}]{2020arXiv200301736H}
{Harrison}, I., {Brown}, M.~L., {Tunbridge}, B., et~al., 2020, arXiv e-prints,
  arXiv:2003.01736, \eprint arXiv:{2003.01736}

\bibitem[{{He} et~al.(2015){He}, {Zhang}, {Ren} \& {Sun}}]{he15}
{He}, K., {Zhang}, X., {Ren}, S., {Sun}, J., 2015, arXiv e-prints,
  arXiv:1512.03385, \eprint arXiv:{1512.03385}

\bibitem[{{Hopkins} et~al.(2002){Hopkins}, {Miller}, {Connolly}, {Genovese},
  {Nichol} \& {Wasserman}}]{2002AJ....123.1086H}
{Hopkins}, A.~M., {Miller}, C.~J., {Connolly}, A.~J., {Genovese}, C., {Nichol},
  R.~C., {Wasserman}, L., 2002, \aj, 123, 2, 1086, \eprint
  arXiv:{astro-ph/0110570}

\bibitem[{{Hopkins} et~al.(2015){Hopkins}, {Whiting}, {Seymour}
  et~al.}]{hopkins2015}
{Hopkins}, A.~M., {Whiting}, M.~T., {Seymour}, N., et~al., 2015, \pasa, 32,
  e037, \eprint arXiv:{1509.03931}

\bibitem[{Hu et~al.(2017)Hu, Shen, Sun \& Albanie}]{articleSE}
Hu, J., Shen, L., Sun, G., Albanie, S., 2017, IEEE Transactions on Pattern
  Analysis and Machine Intelligence, PP

\bibitem[{Ioffe \& Szegedy(2015)}]{Ioffe2015BatchNA}
Ioffe, S., Szegedy, C., 2015, ArXiv, abs/1502.03167

\bibitem[{Jones et~al.(2001)Jones, Oliphant, Peterson et~al.}]{SciPy}
Jones, E., Oliphant, T., Peterson, P., et~al., 2001, {SciPy}: Open source
  scientific tools for {Python}, [Online; accessed ]

\bibitem[{{Kapi{\'n}ska} et~al.(2017){Kapi{\'n}ska}, {Terentev}, {Wong}
  et~al.}]{kapinska17}
{Kapi{\'n}ska}, A.~D., {Terentev}, I., {Wong}, O.~I., et~al., 2017, \aj, 154,
  6, 253, \eprint arXiv:{1711.09611}

\bibitem[{Li \& Tam(1998)}]{LI1998771}
Li, C., Tam, P., 1998, Pattern Recognition Letters, 19, 8, 771 , ISSN 0167-8655

\bibitem[{{Lin} et~al.(2016){Lin}, {Doll{\'a}r}, {Girshick}, {He}, {Hariharan}
  \& {Belongie}}]{lin16}
{Lin}, T.-Y., {Doll{\'a}r}, P., {Girshick}, R., {He}, K., {Hariharan}, B.,
  {Belongie}, S., 2016, arXiv e-prints, arXiv:1612.03144, \eprint
  arXiv:{1612.03144}

\bibitem[{Lin et~al.(2016)Lin, Dollár, Girshick, He, Hariharan \&
  Belongie}]{articleFPN}
Lin, T.-Y., Dollár, P., Girshick, R., He, K., Hariharan, B., Belongie, S.,
  2016

\bibitem[{Liu et~al.(2016)Liu, Anguelov, Erhan et~al.}]{liu2016ssd}
Liu, W., Anguelov, D., Erhan, D., et~al., 2016, in ECCV

\bibitem[{{Lukic} et~al.(2019){Lukic}, {de Gasperin} \&
  {Br{\"u}ggen}}]{2019Galax...8....3L}
{Lukic}, V., {de Gasperin}, F., {Br{\"u}ggen}, M., 2019, Galaxies, 8, 1, 3,
  \eprint arXiv:{1910.03631}

\bibitem[{{McMullin} et~al.(2007){McMullin}, {Waters}, {Schiebel}, {Young} \&
  {Golap}}]{McMullin2007}
{McMullin}, J.~P., {Waters}, B., {Schiebel}, D., {Young}, W., {Golap}, K.,
  2007, {CASA Architecture and Applications}, vol. 376 of \emph{Astronomical
  Society of the Pacific Conference Series}, 127

\bibitem[{{Mohan} \& {Rafferty}(2015)}]{2015ascl.soft02007M}
{Mohan}, N., {Rafferty}, D., 2015, {PyBDSF: Python Blob Detection and Source
  Finder}, \eprint ascl:{1502.007}

\bibitem[{{Molinari} et~al.(2011){Molinari}, {Schisano}, {Faustini},
  {Pestalozzi}, {di Giorgio} \& {Liu}}]{2011A&A...530A.133M}
{Molinari}, S., {Schisano}, E., {Faustini}, F., {Pestalozzi}, M., {di Giorgio},
  A.~M., {Liu}, S., 2011, \aap, 530, A133, \eprint arXiv:{1011.3946}

\bibitem[{Price-Whelan et~al.(2018)Price-Whelan, Sipőcz, Günther
  et~al.}]{astropy:2018}
Price-Whelan, A.~M., Sipőcz, B.~M., Günther, H.~M., et~al., 2018, The
  Astronomical Journal, 156, 3, 123, ISSN 1538-3881

\bibitem[{Redmon \& Farhadi(2018)}]{Redmon2018YOLOv3AI}
Redmon, J., Farhadi, A., 2018, ArXiv, abs/1804.02767

\bibitem[{{Ren} et~al.(2015){Ren}, {He}, {Girshick} \& {Sun}}]{ren15}
{Ren}, S., {He}, K., {Girshick}, R., {Sun}, J., 2015, arXiv e-prints,
  arXiv:1506.01497, \eprint arXiv:{1506.01497}

\bibitem[{Robitaille et~al.(2013)Robitaille, Tollerud, Greenfield
  et~al.}]{astropy:2013}
Robitaille, T.~P., Tollerud, E.~J., Greenfield, P., et~al., 2013, Astronomy \&
  Astrophysics, 558, A33, ISSN 1432-0746

\bibitem[{{Robotham} et~al.(2018){Robotham}, {Davies}, {Driver}
  et~al.}]{Robotham2018}
{Robotham}, A.~S.~G., {Davies}, L.~J.~M., {Driver}, S.~P., et~al., 2018,
  \mnras, 476, 3, 3137, \eprint arXiv:{1802.00937}

\bibitem[{{Rowe} et~al.(2015){Rowe}, {Jarvis}, {Mandelbaum}
  et~al.}]{2015A&C....10..121R}
{Rowe}, B.~T.~P., {Jarvis}, M., {Mandelbaum}, R., et~al., 2015, Astronomy and
  Computing, 10, 121, \eprint arXiv:{1407.7676}

\bibitem[{{Sault} et~al.(1995){Sault}, {Teuben} \& {Wright}}]{sault95}
{Sault}, R.~J., {Teuben}, P.~J., {Wright}, M.~C.~H., 1995, {A Retrospective
  View of MIRIAD}, vol.~77 of \emph{Astronomical Society of the Pacific
  Conference Series}, 433

\bibitem[{Simonyan \& Zisserman(2015)}]{simonyan15}
Simonyan, K., Zisserman, A., 2015, in International Conference on Learning
  Representations

\bibitem[{{Taylor}(2005)}]{2005ASPC..347...29T}
{Taylor}, M.~B., 2005, {TOPCAT \&amp; STIL: Starlink Table/VOTable Processing
  Software}, vol. 347 of \emph{Astronomical Society of the Pacific Conference
  Series}, 29

\bibitem[{Vafaei~Sadr et~al.(2019)Vafaei~Sadr, Vos, Bassett, Hosenie, Oozeer \&
  Lochner}]{vafaei2019deepsource}
Vafaei~Sadr, A., Vos, E.~E., Bassett, B.~A., Hosenie, Z., Oozeer, N., Lochner,
  M., 2019, Monthly Notices of the Royal Astronomical Society, 484, 2, 2793

\bibitem[{Van~der Walt et~al.(2014)Van~der Walt, Sch{\"o}nberger,
  Nunez-Iglesias et~al.}]{van2014scikit}
Van~der Walt, S., Sch{\"o}nberger, J.~L., Nunez-Iglesias, J., et~al., 2014,
  PeerJ, 2, e453

\bibitem[{{Vernstrom} et~al.(2016){Vernstrom}, {Scott}, {Wall}, {Condon},
  {Cotton} \& {Perley}}]{tessa2016}
{Vernstrom}, T., {Scott}, D., {Wall}, J.~V., {Condon}, J.~J., {Cotton}, W.~D.,
  {Perley}, R.~A., 2016, \mnras, 461, 3, 2879, \eprint arXiv:{1603.03084}

\bibitem[{{Virtanen} et~al.(2020){Virtanen}, {Gommers}, {Oliphant}
  et~al.}]{2020SciPy-NMeth}
{Virtanen}, P., {Gommers}, R., {Oliphant}, T.~E., et~al., 2020, Nature Methods,
  17, 261

\bibitem[{{W}es {M}c{K}inney(2010)}]{mckinney-proc-scipy-2010}
{W}es {M}c{K}inney, 2010, in {P}roceedings of the 9th {P}ython in {S}cience
  {C}onference, edited by {S}t\'efan van~der {W}alt, {J}arrod {M}illman, 56 --
  61

\bibitem[{{Whiting} \& {Humphreys}(2012)}]{2012PASA...29..371W}
{Whiting}, M., {Humphreys}, B., 2012, \pasa, 29, 3, 371, \eprint
  arXiv:{1208.2479}

\bibitem[{{Whiting}(2012)}]{2012MNRAS.421.3242W}
{Whiting}, M.~T., 2012, \mnras, 421, 4, 3242, \eprint arXiv:{1201.2710}

\bibitem[{{Wong} et~al.(2020){Wong}, {Garon}, {Alger} et~al.}]{wong20}
{Wong}, O.~I., {Garon}, A., {Alger}, M.~J., et~al., 2020, {Radio Galaxy Zoo
  Data Release 1: quantified visual classificatios of $>$70,000 sources from
  the FIRST and ATLAS surveys}, in preparation

\bibitem[{{Wright} et~al.(2010){Wright}, {Eisenhardt}, {Mainzer}
  et~al.}]{wright10}
{Wright}, E.~L., {Eisenhardt}, P. R.~M., {Mainzer}, A.~K., et~al., 2010, \aj,
  140, 6, 1868, \eprint arXiv:{1008.0031}

\bibitem[{{Wu} et~al.(2019){Wu}, {Wong}, {Rudnick} et~al.}]{wu19}
{Wu}, C., {Wong}, O.~I., {Rudnick}, L., et~al., 2019, \mnras, 482, 1, 1211,
  \eprint arXiv:{1805.12008}

\bibitem[{Zhang et~al.(2017)Zhang, Zuo, Chen, Meng \& Zhang}]{zhang2017beyond}
Zhang, K., Zuo, W., Chen, Y., Meng, D., Zhang, L., 2017, IEEE Transactions on
  Image Processing, 26, 7, 3142

\bibitem[{Zhao et~al.(2019)Zhao, Sheng, Wang et~al.}]{zhao2019m2det}
Zhao, Q., Sheng, T., Wang, Y., et~al., 2019, in Proceedings of the AAAI
  Conference on Artificial Intelligence, vol.~33, 9259--9266

\bibitem[{Zhu et~al.(2019)Zhu, He, Dai, Rao \& Li}]{inproceedingsMSSM}
Zhu, Z., He, M., Dai, Y., Rao, Z., Li, B., 2019, 1789--1794

\end{thebibliography}
\bibliographystyle{mn2e_plus_arxiv}

\appendix

\section{Definition of scores per attribute and per source}\label{sec:wi}

The definition of the $A_{\rm tot}$ and $G_{\rm tot}$ metrics in Sec. \ref{sec:scores} contained weights $w_i$. Each of these weights ranges from 0 to 1 and it quantifies the accuracy of the characterization and classification of the source $i$. In this Appendix we give details on how the weights $w_i$ are computed.

The source properties considered for computing $w_i$ are seven: position (the best between core and centroid position if both are present), flux density, core fraction, major axis, minor axis, position angle and class. For all of them except the source class, the score per source $i$ and per attribute $j$,  $w_i^j$ , is defined as
\begin{equation}
    w_i^j=\frac{1}{7}\min \left\{1,\frac{{\rm D}_i^j}{{\rm thr}^j}\right\}
\end{equation}
where D$_i^j$ is the error on the attribute $j$ for the source $i$ and thr$^j$ is a threshold set on that attribute for all sources. 
 The definition of errors and thresholds for these source properties are in Table \ref{tab:wi}. 
\begin{table}
    \centering
        \caption{Definition of errors and thresholds for source attributes.  \label{tab:wi}}
    \begin{tabular}{c|c|c}
         Attribute&Error&Threshold  \\
\hline
         Position&eq. (\ref{dpos})&0.3\\
         Flux density&eq. (\ref{dflux})&0.1\\
         Major axis&eq. (\ref{dsize})&0.3\\
         Minor axis&eq. (\ref{dsize})&0.3\\
         Position angle&eq. (\ref{dpa})&1\\
         Core fraction&eq. (\ref{dcf})&0.05\\
    \end{tabular}
   
\end{table}

The behaviour of $w_i^j$ is shown in Fig. A1. The maximum score of 1/7 is awarded if the error on the considered attribute is below the threshold; above the threshold the score decreases and tends to 0 for ${\rm D}_i^j \gg {\rm thr}^j$. 

\begin{figure}
    \centering
    \includegraphics[width=8cm]{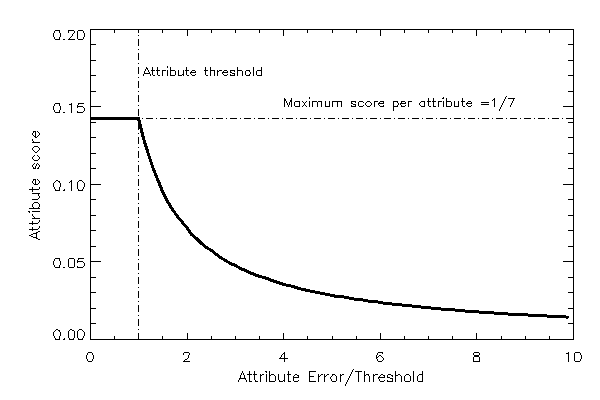}
    \caption{Score awarded per attribute as a function of attribute error/threshold.}
    \label{fig:my_label}
\end{figure}

The thresholds have been calibrated on the global error statistics for the submissions, so that the full range of scores from 0 to 1/7 are awarded on the full sample of results. 
In the case of the source class, $w_i^j$  is either 1/7 or 0 depending on whether the source has been classified correctly or not. The final score per source is finally 
\begin{equation}
w_i=\sum_{j=1}^7w_i^j.
\end{equation}
The 1/7 normalization of $w_i^j$ guarantees that the maximum value of $w_i$ is 1; this is awarded whenever all attributes of that source have been estimated with an error lower than the set threshold and the class has been correctly identified. 

\newpage
\section{JLRAT additional figures}\label{sec:jlrat figs}

\begin{figure*}
\centering
\includegraphics[width=1\textwidth]{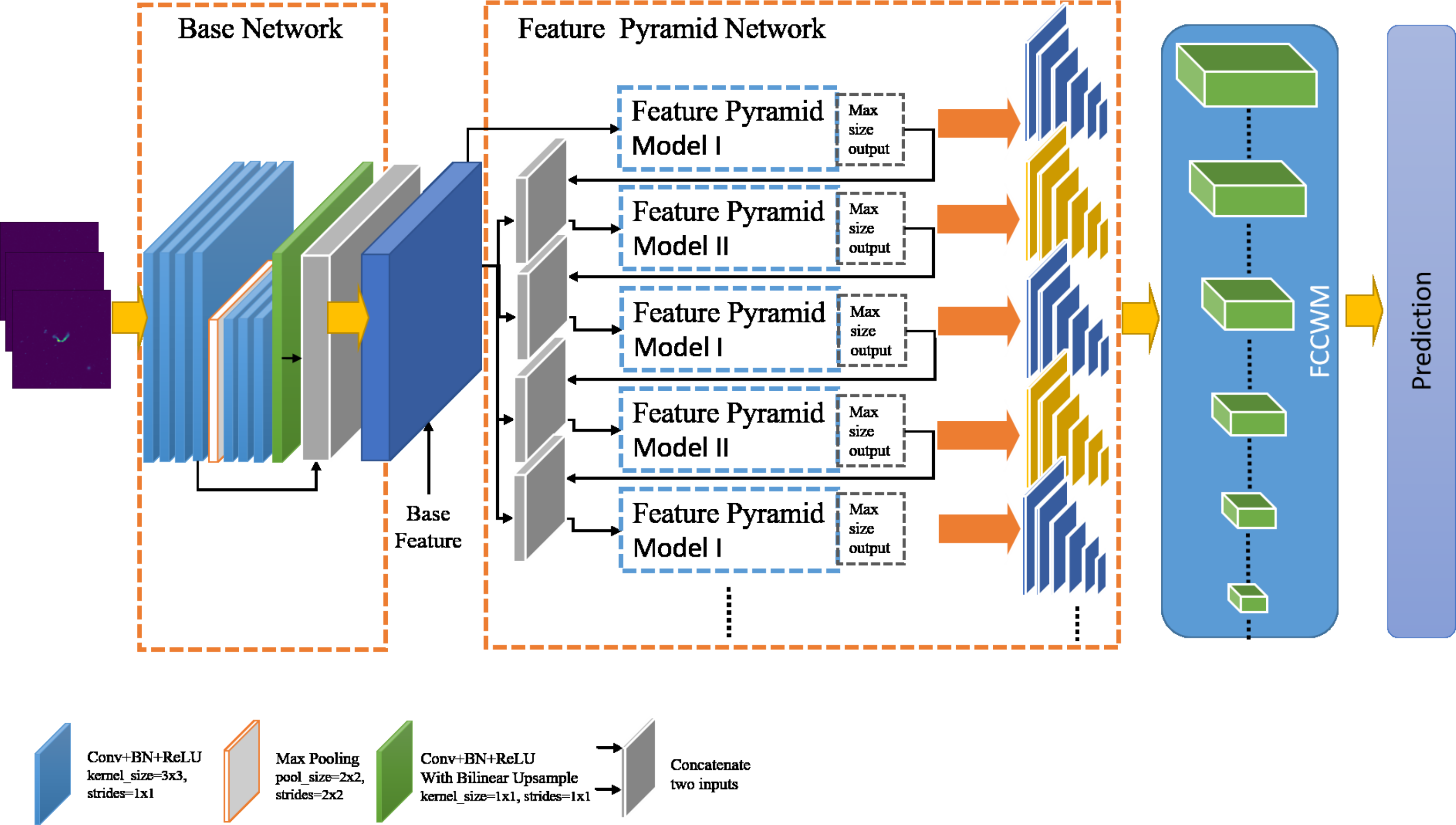}
\includegraphics[width=1\textwidth]{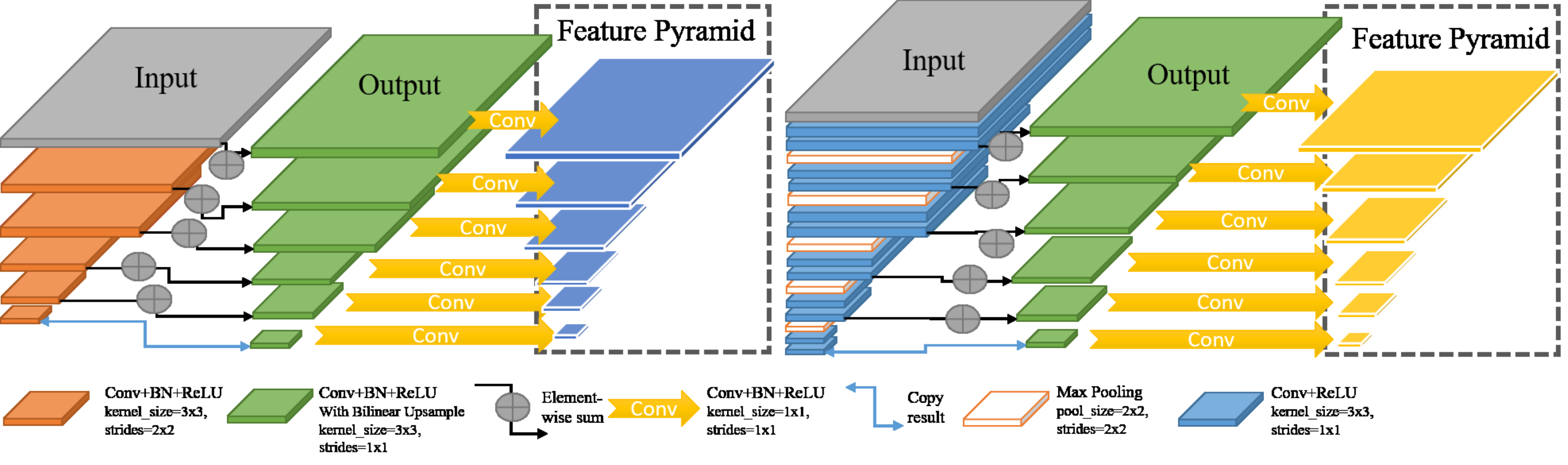}
\caption{\emph{Top:} the Source Detection Model of JSFM. \emph{Bottom:} JSFM feature pyramid models I (left) and II (right).}
\label{fig:m1m2JLRAT}
\end{figure*}

\end{document}